\documentclass[11pt,preprint]{aastex}

\def\spose#1{\hbox to 0pt{#1\hss}}
\def\simlt{\mathrel{\spose{\lower 3pt\hbox{$\mathchar"218$}}
     \raise 2.0pt\hbox{$\mathchar"13C$}}}
\def\simgt{\mathrel{\spose{\lower 3pt\hbox{$\mathchar"218$}}
     \raise 2.0pt\hbox{$\mathchar"13E$}}}
\slugcomment{To be submitted to AJ}
\usepackage{epsf}
\font\smcap=cmcsc10

\shorttitle{Dynamics \& Stellar Content of Giant M31 Stream}
\shortauthors{Guhathakurta et~al.}

\begin{document}

\title{Dynamics and Stellar Content of the Giant Southern Stream in M31.\\
I.~Keck\altaffilmark{1} Spectroscopy of Red Giant Stars}

\author{
Puragra~Guhathakurta\altaffilmark{2},
R.~Michael~Rich\altaffilmark{3},
David~B.~Reitzel\altaffilmark{3},
Michael~C.~Cooper\altaffilmark{4},
Karoline~M.~Gilbert\altaffilmark{2},
Steven~R.~Majewski\altaffilmark{5},
James~C.~Ostheimer\altaffilmark{5},
Marla~C.~Geha\altaffilmark{6,7},
Kathryn~V.~Johnston\altaffilmark{8}, and
Richard~J.~Patterson\altaffilmark{5}
}

\email{
raja@ucolick.org,
rmr@astro.ucla.edu,
reitzel@astro.ucla.edu,
cooper@astron.berkeley.edu,
kgilbert@astro.ucsc.edu,
srm4n@virginia.edu,
jostheim@alumni.virginia.edu,
mgeha@ociw.edu,
kvj@astro.wesleyan.edu,
rjp0i@virginia.edu
}

\altaffiltext{1}{Data presented herein were obtained at the W.\ M.\ Keck
Observatory, which is operated as a scientific partnership among the
California Institute of Technology, the University of California and the
National Aeronautics and Space Administration.  The Observatory was made
possible by the generous financial support of the W.\ M.\ Keck Foundation.}

\altaffiltext{2}{UCO/Lick Observatory, Dept.\ of Astronomy \& Astrophysics,
Univ.\ of California, Santa Cruz, CA 95064}
\altaffiltext{3}{Dept.\ of Physics \& Astronomy, Univ. of California, Los
Angeles, CA 90095}
\altaffiltext{4}{Dept.\ of Astronomy, Univ. of California, Berkeley, CA
94720}
\altaffiltext{5}{Dept.\ of Astronomy, Univ.\ of Virginia, Charlottesville,
VA 22903}
\altaffiltext{6}{Carnegie Observatories, 813 Santa Barbara St., Pasadena, CA
91101}
\altaffiltext{7}{Hubble fellow}
\altaffiltext{8}{Van Vleck Observatory, Wesleyan Univ., Middletown, CT 06459}

\begin{abstract}

This paper presents the first results from a large spectroscopic survey of
red giant branch (RGB) stars in M31 using DEIMOS on the Keck 10-meter
telescope.  A photometric pre-screening method, based on the
intermediate-width $DDO51$ band centered on the Mg\,b/MgH absorption feature,
was used to select spectroscopic targets.  RGB candidates were targeted in a
small section of M31's giant southern tidal stream at a projected distance of
31~kpc from the galaxy's center.  We isolate a clean sample of 68~RGB stars
by removing contaminants (foreground Milky Way dwarf stars and background
galaxies) using a combination of spectroscopic, imaging, and photometric
methods; the surface-gravity-sensitive Na\,{\smcap i} doublet is particularly
useful in this regard.  About 65\% of the M31 stars are found to be members
of the giant southern stream while the rest appear to be members of the
general halo population.  The mean (heliocentric) radial velocity of the
stream in our field is $-458$~km~s$^{-1}$, blueshifted by $-158$~km~s$^{-1}$
relative to M31's systemic velocity, in good agreement with recent velocity
measurements at other points along the stream.  The intrinsic velocity
dispersion of the stream is found to be $15_{-15}^{+8}$~km~s$^{-1}$ (90\%
confidence limits).  A companion paper by \citet{fon04} discusses possible
orbits, implications of the coldness of the stream, and properties of the
progenitor satellite galaxy.  The kinematics, and possibly the metallicity
distribution, of the general halo (i.e.,~non-stream) population in this
region of M31 indicate that it is significantly different from samples drawn
from other parts of the M31 halo; this is probably an indication of
substructure in the halo.  The stream appears to have a higher mean
metallicity than the general halo: $\rm\langle[Fe/H]\rangle\sim-0.54$ versus
$-0.74$, and a smaller metallicity spread.  The relatively high metallicity
of the stream implies that its progenitor must have been a luminous dwarf
galaxy.  The Ca\,{\smcap ii} triplet line strengths of the M31 RGB stars are
generally consistent with photometric estimates of their metallicity (derived
by fitting RGB fiducials in the color-magnitude diagram).  There is indirect
evidence of a population of intermediate-age stars in the stream.

\end{abstract}


\section{Introduction}\label{sec:intro}

The growth of galactic halos through the accretion of smaller stellar
subsystems has been the subject of many studies over the last few decades
\citep{sea78,whi78}.  Numerical simulations and semi-analytical modeling of
the accretion process have reached new levels of detail in recent years
\citep*[e.g.,][]{joh96,joh98,hel99,hel00,bul01}.  The discovery of the
Magellanic Stream \citep*{mat74} provided early observational evidence of an
ongoing accretion/merger event in the Galaxy involving the Large and Small
Magellanic Clouds.  The fact that the Magellanic Stream is seen only in
neutral hydrogen and not stars has led to some debate over whether
ram-pressure stripping or tidal forces are at play
\citep*{moo94,put99,mad02}.  The best example of an ongoing accretion event
in the Milky Way is the Sagittarius dwarf satellite galaxy \citep*{iba94}
with its associated tidal debris \citep{maj03,new03}.  More recently, Sloan
Digital Sky Survey (SDSS) stellar density maps have revealed that the
low-luminosity, remote star cluster Palomar~5 is undergoing tidal disruption
\citep{ode01,roc02}.  SDSS and Two-Micron All-Sky Survey data have led to the
discovery and characterization of the Monoceros stream, an arc-like structure
at low Galactic latitude that is probably the result of an encounter with a
dwarf galaxy \citep{yan03,roc03}.  The Monoceros stream and other Milky Way
structures like it are only just being identified: Their large angular extent
requires the use of wide-field surveys.  Moreover followup studies of these
structures can prove to be difficult from our vantage point within the
Galaxy's disk.

In contrast to the Milky Way, its neighbor, the Andromeda spiral galaxy
(M31), offers certain advantages for halo studies: Its disk is highly
inclined, we have a global external perspective of the galaxy, and yet it is
close enough to allow us to characterize the properties of individual stars
in detail.  \citet{iba01} and \citet{fer02} present star-count maps covering
a large area around M31 and find a giant stream extending to its south
(hereafter referred to as the `giant southern stream') along with several
other signs of disturbance in the halo and outer disk.  While the giant
southern stream appears to be tidal debris from a merger event, the
origin/nature of the other features is not clear.  An investigation of M31's
innermost satellites by \citet*{cho02} confirms the presence of
tidally-distorted outer isophotes in NGC~205 and revealed ongoing stripping
in M32, with the amount of mass lost estimated to be of the same order as
that seen in the stream.  However, any association between the stream and M32
(or any other satellite) based on the above studies is largely
circumstantial at this point.

Spectroscopy of large samples of individual stars in M31 is the only secure
approach to establishing connections between specific streams/features and
satellites, and to global mapping of the kinematics and chemical composition
of the halo.  \citet[][hereafter RG02]{rei02} carried out spectroscopy of
$\sim100$ candidate red giant branch (RGB) stars in an outer minor-axis halo
field in M31 using the Keck 10-meter telescope and Low Resolution Imaging
Spectrograph \citep[LRIS;][]{oke95}.  Another 150~spectra of M31 RGB
candidates in inner minor-axis and outer disk fields are analyzed by
\citet{guh02} and \citet*{rei04}.  Recent papers by \citet[][see their
Figs.~1 and 3]{iba04} and \citet[][see their Fig.~3]{mcc04} present radial
velocities for several dozen RGB stars in M31.  All of these spectroscopic
studies taken together have shed light on a variety of topics including the
dynamics and metallicity distribution of the M31 halo, the relation between
the halo and outer disk, evidence of faint debris trails in the halo, and the
orbit of the giant southern stream.  There is an important question that
remains unanswered though: What fraction of the halo is composed of
identifiable streams?

In Fall 2002 we started a spectroscopic survey of the M31 halo with the new
Deep Imaging Multi-Object Spectrograph \citep[DEIMOS;][]{fab03} on Keck.
This survey has yielded spectra for several hundred RGB candidates to date.
The exposures are deep enough to yield information on both radial velocity
and spectral absorption features; we consider the latter to be especially
important for understanding the formation and evolution of the halo.  The
broad goals of the project are to characterize the dynamics, chemical
abundance distribution, and structure/sub-structure of the M31 halo, with an
emphasis on the global statistics and properties of debris trails from past
mergers.

In this, the first paper from our DEIMOS survey, we focus our attention on
the giant southern stream in M31.  Details of the spectroscopic data set on
which this paper is based are described in \S\,\ref{sec:data}, including
target selection, observations, data reduction and verification methods,
survey efficiency, success rate of the target selection procedure, and
velocity measurement error.  The division of our stellar sample into M31
stream RGB, M31 general halo RGB, and foreground Galactic dwarf populations
is described next in \S\,\ref{sec:sort}.  The dynamics of the stream and halo
are described in \S\,\ref{sec:dyn}.  The metallicity distributions of the
stream and general halo are compared in \S\,\ref{sec:feh} based on broad-band
photometry and spectral absorption features.  The main points of the paper
are summarized in \S\,\ref{sec:summary}.  The companion paper by
\citet[][hereafter Paper~II]{fon04} uses these and other data to obtain
constraints on the orbit of the stream and the nature of the possible
progenitor.

\section{Data}\label{sec:data}

\subsection{Spectroscopic Target Selection}\label{sec:targ_sel}

\subsubsection{Photometric Pre-Screening of M31 Red Giants}\label{sec:ddo51}

Candidate RGB stars in M31's giant southern stream were selected from the
field~`a3' photometry/astrometry catalog of \citet{ost02}.  The catalog is
based on Kitt Peak National Observatory (KPNO)\footnote{Kitt Peak National
Observatory of the National Optical Astronomy Observatories is operated by
the Association of Universities for Research in Astronomy, Inc., under
cooperative agreement with the National Science Foundation.} 4-meter
telescope and MOSAIC camera images in the Washington system $M$ and $T_2$
bands as well as the intermediate-width $DDO51$ band.  Sources were
identified and photometered using the DAOPHOT~II and ALLFRAME software
packages \citep{ste92,ste94}.  The photometric transformation relations of
\citet{maj00}:
$$V\,=\,-0.006+M-0.200\,(M-T_2)$$
\begin{equation}
I\,=\,T_2
\label{eqn:phot_trans}
\end{equation}
are used to obtain magnitudes on the Johnson/Cousins system (the $T_2$ band
is essentially identical to the $I$ band).  The \citet{ost02} survey obtained
KPNO/MOSAIC data in ten~fields around the M31 halo, six of them on the
southeast minor axis extending out to $\gtrsim10^\circ$ from the galaxy
center.  The observations were carried out before the discovery of the giant
southern stream and it is pure chance that field~`a3' happens to intersect
the stream.

The location of the $35'\times35'$ field~`a3' relative to the stream is shown
in Figure~\ref{fig:fld_loc} (bold square).  The field center is located
$\xi=+74.8'$~(east) and $\eta=-127.5'$~(south) with respect to M31's center.
Only a portion of the field, the southwest or lower right half, is on the
stream: The northeast edge of the stream runs more or less diagonally across
field~`a3' from northwest to southeast.  It should be noted that the surface
density of luminous M31 RGB stars is quite low in this remote field, even in
the on-stream portion, and there is substantial foreground and background
contamination (see \S\,\ref{sec:success}).

The $DDO51$ filter has a passband of width $\rm\Delta\lambda\approx100\AA$
centered at $\rm\lambda\sim5150\AA$, designed to include the
surface-gravity-sensitive Mg\,{\smcap i} triplet and MgH stellar absorption
features \citep{maj00}.  The features are strong in dwarf stars but weak in
RGB stars.  Following \citet{pal03} each object is assigned a probability of
being an RGB star, $P_{\rm giant}$, based on the degree of overlap of its
photometric error ellipse with the (pre-determined) locus of dwarf stars in
the ($M-DDO51$) versus ($M-T_2$) color-color diagram.  The $P_{\rm giant}$
parameter is effective in guiding the selection of RGB stars and in reducing
greatly dwarf contamination in the sample, but it is not a perfect
discriminant at the relatively faint magnitudes of our survey (as will be
shown in \S\,\ref{sec:reject_ddo51} below): Metallicity variations cause RGB
stars to have a relatively broad distribution in two-color space and the few
more metal-rich ones that happen to intersect the dwarf locus are assigned a
low $P_{\rm giant}$ value; moreover, RGB (dwarf) stars in the tails of the
photometric error distribution can scatter close to (far from) the dwarf
locus and would then be assigned a low (high) $P_{\rm giant}$ value.  Since
metallicity is a secondary parameter (after surface gravity) in determining
the strength of the Mg\,b and MgH features, one might worry about $DDO51$
selection introducing a bias against metal-rich RGB stars.  We will examine
this issue in \S\,\ref{sec:feh_bias} below and show that such a bias is not
important for our sample.

The object detection algorithm in DAOPHOT tends to reject sources that are
extended relative to the point spread function but some background field
galaxies, especially compact ones, do slip through into the object catalog.
The two-color method described above tends to assign high $P_{\rm giant}$
values to galaxies because they are dominated by the light of RGB stars
and/or because their Mg\,b/MgH features are redshifted out of the $DDO51$
passband making them appear featureless.  For this reason the $P_{\rm giant}$
criterion is supplemented by the DAOPHOT-based morphological criteria {\tt
chi} and {\tt sharp} to reject galaxies.  Naturally, the compact galaxy
rejection efficiency of DAOPHOT's source finding algorithm and {\tt chi}/{\tt
sharp} parameters depends critically on the seeing and depth of the
KPNO/MOSAIC imaging data.

\subsubsection{Slitmask Design}\label{sec:dsim}

Three~DEIMOS multi-slit masks (\#1--\#3) were designed for field~`a3' using
Drew Phillips' {\tt dsimulator} software (see {\tt
http://www.ucolick.org/$\sim$phillips/deimos\_\,ref/masks.html} for
details).  Table~\ref{tab:masks} contains a summary of relevant information
for the masks.  The {\tt dsimulator} software takes as input multiple
lists of spectroscopic targets: lists~1, 2, 3, etc., in order of decreasing
priority.  In addition the targets in each list are assigned weights in
direct proportion to their $P_{\rm giant}$ values.  The user chooses the mask
pointing center and position angle (see thin rectangles in the
Fig.~\ref{fig:fld_loc} inset and Table~\ref{tab:masks}).  For each slitmask
design the software starts with the highest priority list (list~1) and
automatically fills in the $\approx16'\times4'$ mask area to the extent
possible.  It maximizes the sum of weights for the selected targets given
two~constraints: (i)~a minimum length of~$6''$ for each slitlet, with the
target at least $2.5''$ from each end of the slitlet, and a $0.5''$ gap in
the spatial direction between the ends of adjacent slitlets; and
(ii)~avoidance of inter-CCD gaps and vignetted regions, whose locations
within the slitmask are predicted on the basis of an optical model of the
spectrograph.  Next the software fills in available spaces on the slitmask
with targets drawn from the next input list in the priority sequence
(list~2), and so on for lists~3 and~4.

\begin{table}[ht!]
\begin{center}
\caption{Slitmask design parameters and details of observations for the
three~DEIMOS masks that form the basis of this paper.  The number of
spectroscopic science targets selected from lists~1--3 for masks~\#1 and~\#2
and from lists~1--4 for mask~\#3 (in order of decreasing priority; see
\S\,\ref{sec:dsim}) is indicated; this number does not include alignment
stars (list~0).  The number of objects is also broken down according to $Q$,
a code indicating reliability of the measured redshift and spectral quality
(\S\,\ref{sec:zspec}).
}
\vskip 0.3cm
\begin{tabular}{rrrr}
\hline
\hline
                  & Mask~\#1        & Mask~\#2        & Mask~\#3        \\
\hline
Pointing center:~~~
		  &                 &                 &                 \\
$\alpha_{\rm J2000}$ ($\rm^h$:$\rm^m$:$\rm^s$)
                  & 00:48:21.16     & 00:47:47.24     & 00:48:23.17     \\
$\delta_{\rm J2000}$ ($^\circ$:$'$:$''$)
                  & +39:02:39.2     & +39:05:56.3     & +39:12:38.5     \\
Position angle ($^\circ$ E of N)
                  & 64.2            & 178.2           & 270.0           \\
No.\ of spectroscopic science targets
                  & 85              & 80              & 83              \\
Breakdown of targets by lists~0/1/2/3/4
                  & 5/53/23/9/...   & 3/54/15/11/...  & 4/49/17/16/1    \\
Date of observations (UT)
                  & 2002 August 16  & 2002 October 11 & 2003 October 26 \\
Exposure time (s) & $2\times1800$   & $3\times1800$   & $3\times1200$   \\
Number of $Q=-2$/1/2/3/4 cases
                  & 8/22/14/15/26   & 6/19/14/9/32    & 2/0/22/11/48    \\
\hline
\end{tabular}
\label{tab:masks}
\end{center}
\end{table}

The mask~\#1 and~\#2 designs are each based on three~input lists with the
following selection criteria:
\begin{itemize}
\item[{\bf 1.}]{Objects with $20<I<22$, $P_{\rm giant}>0.5$, ${\tt chi}<1.3$,
and $-0.3<{\tt sharp}<+0.3$.}
\item[{\bf 2.}]{Same as list~1 but with the $I$~mag range expanded to $I<20$
(above RGB tip) and $22<I<22.5$ (fainter RGB).}
\item[{\bf 3.}]{Objects with $20<I<22.5$, a less stringent $DDO51$ constraint
($P_{\rm giant}>0.25$), and slightly less stringent {\tt chi}/{\tt sharp}
morphology cuts.}
\end{itemize}

After the initial round of observations in Fall 2002, we decided to make some
changes to the target selection procedure for our survey in order to fill the
slitmasks more efficiently.  The design for mask~\#3 (and the other Fall 2003
masks from our DEIMOS survey that are not presented here) is based on
four~input lists.  The criteria for lists~1 and~2 are the same as those
listed above for the first two~masks.  The list~3 criteria are the same as
those listed above except that the $P_{\rm giant}$ requirement is dropped
altogether.  The list~4 criteria are the same as for the modified list~3 but
with further relaxation of the morphology cuts {\tt chi}/{\tt sharp}.

The number of spectroscopic targets on masks~\#1, \#2, and \#3 is: 85
(53/23/9), 80 (54/15/11), and 83 (49/17/16/1), respectively, where the
numbers in parentheses indicate the breakdown by {\tt dsimulator} input list
number.  Throughout the rest of this paper we collectively (and loosely)
refer to all these targets as ``RGB candidates'' even though many of them
have low $P_{\rm giant}$ values (lists~3 and 4).  The slitlet width is set to
$1''$.  In addition to these science targets, 3--5~bright stars per mask were
selected for the purposes of slitmask alignment using the following (list~0)
criteria: $I<20$, same morphology cuts as lists~1 and 2, and no $P_{\rm
giant}$ requirement.  Each alignment star is assigned a $4''\times4''$ box
while avoiding overlap with the science target slitlets.  After this step
{\tt dsimulator} was used to maximize the lengths of all slitlets in the
spatial direction while maintaining a $0.5''$ inter-slit separation.  A few
guide stars were also selected for each mask using the same list~0 criteria
as for alignment stars; they are useful for coarse mask alignment and guiding
off the TV guider camera, but no spectra are obtained for them.

There are 27~cases of duplication among science targets across the
three~masks: 15~in common between masks~\#1 and~\#2 and 12~in common between
masks~\#2 and~\#3.  No object was observed on all three~masks.  This level of
duplication (15\%--20\%) was achieved by design; in \S\,\ref{sec:vel_error}
we estimate the measurement error in radial velocity from these duplicate
measurements.  The total number of unique M31 RGB candidates targeted across
the three~slitmasks is~221.

Figure~\ref{fig:cmd} shows a color-magnitude diagram (CMD) of objects in the
W portion of field~`a3' from which the DEIMOS spectroscopic targets were
selected.  Extinction/reddening corrections have been applied on a
star-by-star basis using the \citet*{sch98} dust map.  The dust is expected
to be mostly, if not entirely, in our Galaxy (given that field~`a3' is well
removed from M31's disk) so a standard slope of $R_V=3.1$ is assumed for the
extinction law which translates to $E(V-I)/E(B-V)=1.4$ \citep*{car89}.
Panel~({\it a\/}) shows all objects, without any photometric or morphological
screening; panel~({\it b\/}) shows only those objects that pass the $P_{\rm
giant}$ and morphology cuts for lists~1 and~2.  Foreground Galactic dwarf
stars in ({\it a\/}) form a broad swath over the color range
$0.6\lesssim(V-I)_0\lesssim3$, but the cuts appear to be effective at
removing this population at bright magnitudes ($I\lesssim21$).

\subsection{Observations}\label{sec:obs}

Masks~\#1, \#2, and \#3 were observed on 2002 August~16, 2002 October~11, and
2003 October~26 (UT), respectively, using the Keck~II 10-meter telescope and
DEIMOS with its 1200~lines~mm$^{-1}$ grating.  The exposure times were
$2\times1800$~s, $3\times1800$~s, and $3\times1200$~s, respectively.  The
observational details are given in Table~\ref{tab:masks}.  The seeing FWHM
was in the range $0.7''$--$0.9''$.  Standard quartz and arc lamp exposures
were obtained through each mask for rectification, flat fielding, and
wavelength calibration purposes.  For details of the instrument we refer the
reader to \citet{fab03} and {\tt
http://alamoana.keck.hawaii.edu/inst/deimos/}.

The mask~\#1 observation was carried out within two~months of the initial
commissioning of the DEIMOS spectrograph and even the mask~\#2 observation
was carried out during a period when improvements were continually being made
to the instrument; the data are nevertheless of superb quality.  The 28~best
stellar spectra, the top third of our 84-star sample in terms of S/N, are
presented in Figure~\ref{fig:spec_indiv}.  The main spectral feature of
interest to us is the Ca\,{\smcap ii} triplet at $\rm\lambda\sim8500\AA$,
although useful information can also be gleaned from other weaker features in
this `far-red' region of the spectrum (\S\,\ref{sec:feh_spec}).

The central wavelength setting was $\rm\lambda8550\AA$ for masks~\#1 and \#2.
This was changed to $\rm\lambda7800\AA$ for mask~\#3 and the rest of the Fall
2003 masks from our DEIMOS survey (not presented here) to avoid the
possibility of losing one of the Ca\,{\smcap ii} triplet lines in the
inter-CCD gap and to extend the blue side coverage to the $\rm\lambda7100\AA$
TiO and $\rm\lambda6563\AA$ H$\alpha$ features.  The 1200~lines~mm$^{-1}$
grating yields a dispersion of $\rm0.33\,\AA$~pix$^{-1}$ and the spatial
scale is $0.12''$~pix$^{-1}$.  The $4\times2$ array of $\rm2K\times4K$ CCDs
span a spectral range of about $\rm\Delta\lambda\gtrsim2700\AA$ and a total
slitmask length of over $16'$.  Thus, spectra from masks~\#1 and \#2
(Fall~2002) cover the range $\lambda\lambda7200$--$\rm9900\AA$, while
mask~\#3 spectra (Fall~2003) cover $\lambda\lambda6450$--$\rm9150\AA$.  The
width of the mask (in the dispersion direction) over which slitlets are
distributed is about 2000~pix so the exact spectral coverage varies from
slitlet to slitlet by up to a few hundred~\AA; moreover, some spectra are
truncated by vignetting (see Fig.~\ref{fig:spec_indiv}).

The $1''$ slitlet width used on our masks subtends 4.8~pix, given an
anamorphic demagnification factor of~0.57 for the 1200~lines~mm$^{-1}$
grating at $\rm\lambda8500\AA$.  The actual resolution is slightly better
than this: For typical seeing of $0.8''$ (FWHM), the spectral resolution is
$\rm3.8~pix=1.26\AA$ which corresponds to 44~km~s$^{-1}$ or $R\lesssim7000$
at the Ca\,{\smcap ii} triplet.  While this is the characteristic width of
stellar absorption lines, the precision with which the centroid of a line can
be determined is typically a small fraction of its width.  The centroiding
accuracy, and hence the radial velocity measurement error
(\S\,\ref{sec:vel_error}), depends on the S/N ratio or, more specifically, on
the significance of the cross-correlation peak \citep{ton79}.

Prior to starting spectroscopic exposures on each slitmask, we used the mask
alignment procedure developed by the DEEP2 team: Guide stars on the TV guider
camera were used for coarse alignment; direct images were then obtained
through the mask with the grating in zeroth order to fine tune the alignment
(both position angle and translation).  The procedure converged after
2--3~iterations with typical residuals of $\gtrsim0.1''$ between the position
of the alignment star and the center of its alignment box.  This is
indicative of the level of astrometric precision in the \citet{ost02} catalog
from which our spectroscopic targets were drawn.

\subsection{Data Reduction}\label{sec:data_red}

\subsubsection{Pipeline Processing}\label{sec:pipeline}

The three~DEIMOS masks in field~`a3' were processed through the {\tt spec2d}
software pipeline (version~1.1.4) developed by the DEEP2 team at the
University of California-Berkeley (UCB) for that survey.  A brief listing of
the main processing steps is given below; details of the data processing
steps may be found at {\tt
http://astron.berkeley.edu/$\sim$cooper/deep/spec2d/primer.html}.

The flat field exposures are used to rectify the curved spectra in the raw
spectrogram into rectangular arrays by applying shifts in the spatial
direction.  Next, a one-dimensional (1D) slit function correction and
two-dimensional (2D) flat-field and fringing correction are applied to each
slitlet.  Using the DEIMOS optical model as a starting point, a 2D wavelength
solution is determined from the arc lamp exposures with residuals of order
$\rm0.01\AA$.  Each slitlet is then 2D sky-subtracted exposure by exposure
using a B-spline model for the sky.  The individual exposures of the slitlet
are then averaged with cosmic-ray rejection and inverse-variance weighting.
Finally 1D spectra are extracted for all science targets using the optimal
scheme of \citet{hor86} and converted to plain text format using standard
IRAF\footnote{IRAF is distributed by the National Optical Astronomy
Observatories, which are operated by the Association of Universities for
Research in Astronomy, Inc., under cooperative agreement with the National
Science Foundation.} tasks.

The 1D spectra shown in Figure~\ref{fig:spec_indiv} represent the top
one-third of our stellar sample in terms of~S/N: typically
$\rm\approx15\,\AA^{-1}$ at the Ca\,{\smcap ii} triplet for this subset.  The
spectra have been normalized to unit flux at $\rm\lambda8500\AA$; they are
not flux calibrated and the telluric A-band feature is still present (it will
be removed in future versions of the data processing pipeline).  Boxcar
smoothing, with a 10~pix ($\rm3\AA$) window and inverse-variance weighting,
has been applied to the spectra for illustration purposes.

The DEIMOS spectra presented here are generally of higher quality than the
LRIS spectra in the RG02 study, even though the latter typically had
$3\times$ longer exposure times (compare Fig.~\ref{fig:spec_indiv} to their
Fig.~1).  The most significant improvement is Poisson-limited subtraction of
the bright night sky emission lines that plague the Ca\,{\smcap ii} triplet
region.  Improved sky subtraction is the result of three~features of DEIMOS:
(i)~a closed-loop flexure compensation system; (ii)~CCDs that lack fringing
in the far red; and (iii)~a fine enough pixel scale to sample the line spread
function well, including the sharp slit edges.  Moreover DEIMOS has a higher
net throughput than LRIS in this wavelength regime and its spectral coverage
is wider than the red side of LRIS.  In addition to these improvements in
quality, DEIMOS' roughly $2\times$ larger slitmask area allows a higher
degree of multiplexing than LRIS.

The extracted 1D spectra were processed through the {\tt spec1d} pipeline
developed for the DEEP2 survey at UCB (an adaptation of the corresponding
SDSS pipeline).  The pipeline cross-correlates the spectrum of each science
target against a series of stellar templates spanning a range of spectral
types and emission- and absorption-line galaxy templates to determine the
redshift.  The science and template spectra are continuum-subtracted and the
science spectrum is interpolated to the resolution of the template:
$\rm\Delta\log\lambda=2\times10^{-5}\,pix^{-1}$.  The cross-correlation is
computed in pixel space (i.e.,~real space as opposed to Fourier space) with
the relative line strengths and line widths held fixed.  The software shifts
and scales the template to find the best fit in reduced-$\chi^2$ space.  The
galaxy templates used in the fitting procedure are linear combinations of the
emission- and absorption-line templates whereas the various stellar templates
are used individually in the fit.  The 10~best solutions for the redshift $z$
of each object are reported, arranged in order of increasing
reduced-$\chi^2$.

\subsubsection{Quality Assessment}\label{sec:zspec}

We used the visual inspection software {\tt zspec}, developed by D.~Madgwick
et~al.\ at UCB for the DEEP2 survey, to view the sky-subtracted 2D and 1D
spectra of each slitlet/science target.  The extraction window used by the
{\tt spec2d} pipeline---i.e.,~the range of ``rows'' in the 2D spectrum that
are collapsed to form the 1D spectrum---is indicated by markers along the
spatial axis of {\tt zspec}'s 2D spectrum display.  In rare cases this window
appeared to be too narrow/wide or displaced from the target's stellar
continuum (or emission lines); we manually set the extraction window in these
cases, re-extracted the 1D spectrum, and processed it through {\tt spec1d}
pipeline.

The 10~best redshift choices from {\tt spec1d} are listed by {\tt zspec}.
Selecting one of the $z$ choices causes the corresponding (appropriately
redshifted) template to be displayed overlaid on the science target's 1D
spectrum.  The positions of prominent absorption or emission lines in the
template in question (e.g.,~Ca\,{\smcap ii} triplet lines for a stellar
template and [O\,{\smcap ii}], [O\,{\smcap iii}], H$\beta$, etc.\ lines for
an emission-line galaxy template) are marked on the 1D and 2D spectra.  The
night sky spectrum, or more precisely the variance versus wavelength, is
plotted along side the target's 1D spectrum; it proved to be very useful in
deciding which spectral features were reliable and which ones were not.
Different degrees of smoothing were tried on the target's 1D spectrum to
enhance the S/N at the cost of spectral resolution---this afforded a better
view of its spectral features and allowed us to assess the reality of
marginal/weak ones.

For about half the targets from masks~\#1 and~\#2 and two-thirds of those
from mask~\#3 it was easy to pick out the correct $z$ value from the choices
provided by {\tt spec1d}.  On rare occasions, none of the 10~choices was
accurate but it was obvious from the 1D spectrum what the correct $z$ was.
The redshift was marked {\it manually\/} in these cases.  It was typically
based on the Ca\,{\smcap ii} triplet, often a subset of the three~lines (the
reddest line lies amidst a cluster of night sky lines and is sometimes
affected by them), the $\rm\lambda7100\AA$ TiO band (for red stars), and/or
the Na\,{\smcap i} doublet (dwarf stars).  Only 19 out of the 248~science
targets, or 17 out of the 221~unique M31 RGB candidates (8\%), are in this
``$z$-by-hand'' category.

Following the classification scheme used in the DEEP2 survey \citep{coi04},
each science target was assigned a quality code $Q$ to indicate the
reliability of the measured redshift and the overall quality of its spectrum.
The above cases with well-measured $z$'s, including the ``$z$-by-hand''
cases, were placed in one of two~categories: (i)~$Q=4$ for redshifts based on
two or more robust spectral features; and (ii)~$Q=3$ for those based on
one~robust feature and one or more marginal ones or on a few marginal
features.  The $Q=4$ $z$ measurements are expected to be ``rock solid'' with
something like 99\% confidence, whereas the $Q=3$ $z$'s are expected to have
$\gtrsim90\%$ reliability.  No distinction is made between $Q=3$ versus $Q=4$
cases in the rest of this paper; both are treated as secure redshifts.

Objects for which the redshift measurement failed fall into three~categories
designated: (i)~$Q=-2$ to indicate a catastrophic failure in the data
reduction for instrumental reasons such as severe vignetting, the spectrum
landing in the inter-CCD gaps, near the periphery of the CCD array, or on a
bad column, poor sky subtraction, scattered light problems, etc.; (ii)~$Q=1$
for slitlets with a barely visible spectral continuum and/or very low~S/N;
and (iii)~$Q=2$ for cases where the S/N is marginal to adequate but there is
not enough information for reliable $z$ determination.  Judging from the
DEEP2 survey, some of our $Q=2$ cases are probably distant red galaxies for
which the limited spectral coverage and low~S/N prevent us from making a
reliable redshift measurement.  The low~S/N in some of the $Q=1$ cases in our
sample, especially the handful of bright ones [Fig.~\ref{fig:cmd}({\it
c\/})], could be the result of early mask design/fabrication problems.  Such
instrumental failures rightfully belong in the $Q=-2$ category, but we have
not attempted to reclassify any $Q=1$ cases as there is no direct way to
confirm this hypothesis.

A heliocentric correction is applied to each of the three~masks using the
IRAF task RVCOR.  The observation dates span a wide range and the corrections
span a range as well: $+23$~km~s$^{-1}$ (mask~\#1), $+4$~km~s$^{-1}$
(mask~\#2), and $-3$~km~s$^{-1}$ (mask~\#3).

\subsection{Efficiency / Success Rate}\label{sec:success}

In this section we assess the efficiency of our spectroscopic survey and
success rate of the photometric target selection procedure---i.e.,~the yield
of M31 RGB stars.  We do this by counting up the numbers of objects from all
three~masks in the different redshift quality ($Q$) code categories.

Starting with the failed $z$ measurements, the number of $Q=-2/1/2$ cases is
8/22/14, 6/19/14, and 2/0/22 for masks~\#1, \#2, and \#3, respectively
(Table~\ref{tab:masks}).  Most of the 16~catastrophic failures ($Q=-2$) are
near the ends of the masks and there are as many as 7~duplicates among them;
this high duplicate fraction (nearly 50\%) is attributable to the fact that
the overlap between masks occurs only near their ends
(Fig.~\ref{fig:fld_loc}).

The fraction of catastrophic failures is lower for the Fall 2003 mask~\#3
than for the Fall 2002 masks~\#1 and \#2, and the fraction of spectroscopic
successes higher.  There are a few reasons for this.  The Fall 2002 mask
designs were based on a preliminary optical model for the spectrograph in
{\tt dsimulator}: Design imperfections resulted in some slitlets on masks~\#1
and \#2 being affected by vignetting and inter-CCD gaps.  The DEIMOS optical
model was refined in time for the Fall 2003 mask designs.  The procedure for
taking proper calibration data to achieve good wavelength calibration and
flat-fielding was fine-tuned over the course of a few months by the DEEP2 and
DEIMOS teams, so the calibration is less than ideal for the two~early masks.
The Fall 2002 wavelength setting was not optimal in that there was a finite
chance of one of the three Ca\,{\smcap ii} triplet lines landing in the
inter-CCD gap; the wavelength setting was improved for the Fall 2003
observations.  Finally, the observing conditions were slightly sub par for
the two~early masks with occasional thin cirrus and/or worse-than-average
seeing.

It is noteworthy that there is no significant difference in success rate
between masks~\#1 and \#2 even though the integration time is 50\% longer for
the latter.  This suggests that factors other than total exposure time
(e.g.,~such as the factors listed above) are responsible for determining the
final data quality and success rate for the two~early masks.

There are 9~catastrophic failures ($Q=-2$) among the 221~unique RGB
candidates targeted on the three~masks.  While these failures lower the
overall efficiency of our survey, it is important to keep in mind that they
are {\it instrumental\/} failures that reflect in no way on the physical
properties of the targeted objects or on the success rate of the photometric
screening procedure---i.e.,~it is as though these objects were not observed
at all.  For this reason, the percentages quoted below are measured relative
to a denominator of 212~objects ($=221-9$), those for which spectra were
successfully obtained.

The number (percentage) of $Q=1$ and 2 cases is 38 (18\%) and 46 (22\%),
respectively.  Among the remaining 128~objects (60\%) with definite
redshifts, there are 44~galaxies (21\%), mostly compact emission-line
galaxies, spanning the redshift range $z=0.1$--1.5.  That leaves 84~stars
(39\%), of which 68 (32\%) are M31 RGB stars and 16 (7.5\%) are foreground
Galactic dwarf stars, as we will show in \S\,\ref{sec:reject}.  The
efficiency and success rate for the rest of our DEIMOS survey are expected to
be higher than the fractions quoted here: Most of the remaining slitmasks
(i.e.,~those not presented in this paper) were observed in Fall 2003 and
should therefore be comparable to mask~\#3.

It is instructive to examine the foreground/background contamination rates in
the present study in the context of the earlier RG02 survey.  The surface
density of M31 stars at the $R=31$~kpc location of our field~`a3' (including
the contribution of the giant southern stream) should be roughly similar to
that in RG02's $R=19$~kpc minor-axis field; the surface density of foreground
Galactic stars and background field galaxies is also expected to be the same
in the two~fields.  Milky Way dwarf stars constituted an estimated 57\% of
RG02's sample of 80~stars, whereas they represent only 16 out of the 84~stars
(19\%) in the present study, or 6 out of 62~stars (10\%) if one excludes the
``filler'' targets (lists~3 and~4) for which the $P_{\rm giant}$ criterion
was relaxed or dropped.  The suppression of foreground contaminants in our
study is attributable to $DDO51$-based pre-selection of spectroscopic targets
(\S\,\ref{sec:targ_sel}).  Spectroscopically-confirmed galaxies comprise
about a fifth of our sample of 212~targets, but the true galaxy contamination
rate could easily be higher by a factor of~two since some (unknown) fraction
of the $Q=2$ cases (and possibly $Q=1$ cases) are probably background
galaxies.  The efficiency with which the morphological criteria {\tt
chi}/{\tt sharp} reject compact galaxies is very sensitive to the seeing and
depth of the KPNO/MOSAIC image; there is a fair bit of variation in
seeing/depth across the different fields in the \citet{ost02} survey.  The
RG02 study had a lower galaxy contamination rate (estimated to $\sim10$\%)
thanks to their use of four-band photometry to pre-screen against galaxies
and better seeing.

\subsection{Velocity Measurement Error}\label{sec:vel_error}

The 27~duplicate measurements across the three~masks include 13~cases where
both members of the pair are secure $z$ determinations ($Q=3$ or~4): 8~stars,
3~emission-line galaxies, and 2~absorption-line galaxies.  The rms radial
velocity difference between pairs of measurements for the 8~stars is
$21\pm5$~km~s$^{-1}$.  The $\pm1\sigma$ uncertainty in the rms estimate was
derived from 1000~sets of 8~Monte Carlo drawings from a Gaussian
distribution.  Assuming the measurement uncertainty is the same for each
member of the pair, the radial velocity error for an individual measurement
is $\sqrt{2}$ times smaller than the rms of the difference, or
$15\pm3.5$~km~s$^{-1}$.  The velocity error is probably smaller for mask~\#3
than for the other two~masks, but this is ignored in the above estimate.  We
will use this velocity error estimate in \S\,\ref{sec:vel_disp} to constrain
the coldness of the stream.

The overlap areas between masks, where we have duplicate observations, are
located near the ends of the masks (Fig.~\ref{fig:fld_loc}) and the data
quality appears to be sub-optimal in these regions (\S\,\ref{sec:success}).
This provides a plausible explanation for the fact that 5 of the 16~duplicate
star measurements (31\%) are ``$z$-by-hand'' cases, whereas only 13\% of all
secure $z$ determinations are in this category.  The velocity measurement
error for the typical star in our sample is likely to be somewhat smaller
than the above estimate of 15~km~s$^{-1}$.

\section{Sorting Out the Stellar Sample}\label{sec:sort}

In this section, we describe how the 84~confirmed stars in our sample are
sorted into three~groups: (i)~RGB stars in M31's giant southern stream;
(ii)~RGB stars belonging to M31's general halo population; and
(iii)~foreground Galactic dwarf stars.  The surface density of luminous RGB
stars is relatively low at the location of our field~`a3', in the outer halo
of M31 at a projected distance of $R=31$~kpc from the center of the galaxy
close to its minor axis.  Despite the $\approx3\times$ overdensity due to the
presence of the stream (\S\,\ref{sec:contrast}) and the use of $DDO51$
photometry to screen out Milky Way dwarf stars (\S\,\ref{sec:targ_sel}), our
spectroscopic sample contains a non-negligible fraction of foreground dwarf
star (and background field galaxy) contaminants.  We demonstrate below that a
few key pieces of photometric and spectroscopic information can be used to
distinguish between stream and general halo populations and to eliminate all
contaminants from our field~`a3' sample without any siginificant loss of M31
RGB stars.  The radial velocity distribution of stars is a logical starting
point for this analysis.

\subsection{Line-of-Sight Velocity Distribution}\label{sec:losvd}

Figure~\ref{fig:vhist} shows radial velocity histograms for stars in our
field~`a3' spectroscopic sample, first slitmask by slitmask and then for the
combined sample of 84~unique objects.  Three~features of the distribution are
worthy of note: (i)~a strong and narrow peak that dominates each of the
four~histograms and presumably corresponds to M31's giant southern stream;
(ii)~a low-level broad component that is shifted towards less negative
velocities than the main peak ($-500\lesssim{v}<-200$~km~s$^{-1}$); and
(iii)~a weak concentration of stars seen as a disjoint group in the range
$-150\lesssim{v}<0$~km~s$^{-1}$.  A two-Gaussian maximum likelihood fit to
the first two~components is shown as a thin solid line in panel~({\it d\/}).
We will show below that the second~component represents M31's general halo
population (\S\,\ref{sec:halo_dyn}) while the third consists of Galactic
dwarf stars along the line of sight, but well in front of M31 of course
(\S\,\ref{sec:reject}).

Mask~\#3 has the highest success rate of the three~masks for reasons
discussed in \S\,\ref{sec:success}: Their velocity histograms contain 26, 27,
and 39~stars, respectively, even though the number of spectroscopic targets
on the three~masks is comparable.  The foreground dwarf star fraction is
highest in the mask~\#3 velocity histogram because the $P_{\rm giant}$
criterion was dropped entirely while selecting lists~3 and~4 ``filler''
spectroscopic targets (see \S\,\ref{sec:dsim}).

\subsection{Contrast of Stream Against Smooth Halo}\label{sec:contrast}

The two-Gaussian fit indicates that the ratio of giant southern stream to
general halo stars is about 45:23---i.e.,~that the stream comprises
$65_{-21}^{+12}\%$ (90\% confidence limits from the maximum likelihood
analysis) of the M31 population in field~`a3'.  The actual ratio of stream to
general halo stars may be somewhat higher than this: The $I<22.5$ limiting
magnitude, and possibly the $DDO51$ criterion, used in the spectroscopic
target selection process tends to bias the sample against the most metal-rich
RGB stars and these stars constitute a larger fraction of the stream than the
general halo population (\S\,\ref{sec:feh}).

It is important to note that we can only distinguish between stream and
general halo stars on a statistical basis.  The radial velocity range over
which the stream dominates, $v<-410$~km~s$^{-1}$, contains 47~stars of which
45 are estimated to be members of the stream while the remaining two are
members of the halo.  The 21~M31 RGB stars outside this velocity range are
all likely to be members of the halo.

The surface density of the stream appears to be roughly constant along its
length over a wide range \citep[see star count map in][]{fer02}.  Since the
halo density falls monotonically with increasing distance from the galaxy
center, the contrast of the stream is expected to become progressively
stronger.  This is evident from the velocity histograms presented in the
\citet{iba04} study (compare fields~1, 2, 6, and~8 in their Fig.~1); in fact,
the contrast in their innermost field (field~8) is so low that the stream is
not readily discernible as a distinct population.  It should be noted that
the \citeauthor{iba04} fields were chosen to run along the highest surface
density part of stream whereas our field~`a3' sample is drawn from the edge
of the stream (by chance), so the stream-to-general halo ratio in their
fields should be slightly higher than in ours at the same radial distance
from M31.

\subsection{Rejecting Foreground Galactic Dwarf Stars}\label{sec:reject}

This section discusses foreground Milky Way dwarf star contaminants in our
sample.  Galaxies with high quality spectra ($Q=3$ or~4) are easily
identified on the basis of redshift and spectral characteristics and removed
from the sample, but it is a little more difficult to screen out foreground
dwarf stars.  Five~pieces of information can be used to distinguish M31 RGB
stars from foreground Galactic dwarfs: (i)~radial velocity, (ii)~location
within the ($V-I,~I$) CMD, (iii)~$P_{\rm giant}$ parameter derived from the
($M-DDO51$) versus ($M-T_2$) two-color diagram, (iv)~strength of the
$\rm\lambda8190\AA$ Na\,{\smcap i} doublet in red (cool) stars, and
(v)~comparison of photometric versus spectroscopic metallicity estimates.  In
general, no single criterion of the five is by itself a perfect discriminant
between RGB and dwarf stars, but the combination is very effective.  We
discuss each of these criteria in turn below.

\subsubsection{Radial Velocity}\label{sec:reject_rv}

The radial velocity histogram of the combined stellar sample
[Fig.~\ref{fig:vhist}({\it d\/})] has a distinct {\it gap\/} separating a
group of~68~stars with $v<-200$~km~s$^{-1}$ from a group of~16~stars with
$v>-150$~km~s$^{-1}$ (shaded histogram).  The former group of stars includes
the prominent peak associated with the giant southern stream and is roughly
centered on M31's systemic velocity of $-300$~km~s$^{-1}$; they are therefore
designated `candidate M31 RGB stars'.  The latter group of stars occupies the
radial velocity range predicted by the IASG Galactic star-count model
\citep[RG02;][]{bah84,rat85}; they are therefore designated `candidate
Galactic dwarf stars'.  We will now test the validity of these designations
by comparing the properties of these two~groups of stars.

\subsubsection{Color-Magnitude Diagram}\label{sec:reject_cmd}

The CMD locations of the two~subgroups of stars are consistent with the above
designations.  Candidate M31 RGB stars have a distribution that is nicely
bracketed by the model RGB tracks [Fig.~\ref{fig:cmd}({\it d\/})] with the
majority lying below the RGB tip (the few exceptions will be discussed in
\S\,\ref{sec:feh_comp}).  By contrast very few candidate Galactic dwarfs lie
below the RGB tip [Fig.~\ref{fig:cmd}({\it c\/})]---this is not surprising
because the density of Milky Way (thin and thick disk) stars is on the
decline at these faint apparent magnitudes and the probability of including a
foreground dwarf star is further diminished by the onset of M31's RGB [see
Figs.~\ref{fig:cmd}({\it a\/}--{\it b\/})].

\subsubsection{DDO51-Based Selection}\label{sec:reject_ddo51}

A feature of our DEIMOS spectroscopic survey that sets it apart from previous
studies of M31 is $DDO51$-based pre-screening of RGB stars
(\S\,\ref{sec:targ_sel}).  Figure~\ref{fig:vel_gprob} shows a plot of $P_{\rm
giant}$ versus radial velocity for the 84~stars in our sample.  As expected
the majority of candidate Galactic dwarfs (10 out of~16) have $P_{\rm
giant}<0.5$, even though only a small fraction of the total number of
spectroscopic targets are below this cut: only 37~``filler'' targets
(lists~3 and~4) out of a total of 248~targets on the three~masks or 15\%
(Table~\ref{tab:masks}).  By contrast most candidate M31 RGB stars are above
the cut (62 out of~68, or 90\%).  Put another way, the ratio of candidate M31
RGB to Galactic dwarf stars is 62:6 for $P_{\rm giant}>0.5$, and 6:10 for
$P_{\rm giant}<0.5$.  This demonstrates the usefulness of $DDO51$-based
pre-screening for improving the yield of RGB stars.  However, while candidate
M31 RGB stars have a higher mean/median $P_{\rm giant}$ value than candidate
Galactic dwarfs (consistent with their designations), the distributions are
broad and overlap each other so that there is not a clear separation between
the two~populations.

\subsubsection{Sodium Doublet Line Strength}\label{sec:reject_Na_doublet}

\citet{spi69} first developed the use of the $\rm\lambda\lambda8183,~8195\AA$
Na\,{\smcap i} absorption line doublet as a surface-gravity indicator.  More
recently \citet{sch97} have shown that the Na\,{\smcap i} doublet is expected
to be strong in dwarf stars cooler than $T_{\rm eff}\sim4000\,$K, which
corresponds to $(V-I)_0>1.8$ according to the calibration relation of
\citet*{alo99}, while RGB stars and hotter dwarfs are expected to have weak
lines.  We test this diagnostic on the sample of 84~stars in field~`a3'.

Figure~\ref{fig:na_VI0} is a plot of the Na\,{\smcap i} doublet equivalent
width EW(Na) as a function of the dereddened broadband color $(V-I)_0$ for
our stellar sample.  The EW(Na) computation is done within a window of width
$\rm\Delta\lambda=21\AA$ centered on $\rm\lambda8190\AA$ with continuum bands
of roughly twice/half the width on the blue/red sides; this is a departure
from the Na index definition of \citet{sch97}, which is inevitable since our
spectra have much lower resolution than theirs, but should be fine for an
assessment of the {\it relative\/} line strengths of RGB versus dwarf stars.
The error $\rm\sigma[EW(Na)]$ is assumed to scale inversely with S/N and is
estimated empirically from the 8~stars with duplicate measurements.  Red
stars display a bimodal distribution of Na\,{\smcap i} line strengths as
expected.  It is very reassuring that our velocity-based subsamples track
this bimodality.  The high S/N cases with $(V-I)_0>1.8$ in
Figure~\ref{fig:na_VI0} best demonstrate this: Candidate M31 RGB stars
(circles) lie at or below the $\rm EW(Na)=1.8\AA$ threshold (dashed
horizontal line) while candidate Galactic dwarfs (crosses) lie at or above
it.

\subsubsection{Calcium Triplet Line Strength}\label{sec:reject_Ca_triplet}

The final criterion for distinguishing candidate M31 RGB stars from candidate
foreground Galactic dwarfs is a comparison between photometric and
spectroscopic metallicity estimates.  The $\rm[Fe/H]_{phot}$ estimate is
derived from the position of the star in the CMD relative to model RGB tracks
(see \S\,\ref{sec:feh_phot_method} below).  The $\rm[Fe/H]_{spec}$ estimate
is derived from the strength of the Ca\,{\smcap ii} absorption line triplet
and empirical calibration relations based on RGB stars in Galactic globular
clusters (\S\,\ref{sec:sigma_ca_pred}).  RG02 found that M31 RGB stars lie
close to the $\rm[Fe/H]_{spec}=[Fe/H]_{phot}$ line, whereas Galactic dwarfs
have weaker Ca\,{\smcap ii} lines so that $\rm[Fe/H]_{spec}<[Fe/H]_{phot}$
for the most part (see their Figs.~13 and~14).  We will show in
Figure~\ref{fig:ca_pred_meas} and \S\,\ref{sec:sigma_ca_comp} below that the
Ca\,{\smcap ii} triplet line strengths measured in coadded spectra of M31 RGB
stars and Galactic dwarfs in our field~`a3' sample follow these same trends.

\subsubsection{Discussion: Towards a Clean and Complete Sample of M31 Red
Giant Stars}\label{sec:reject_clean}

Of the five~diagnostics discussed above, the Na\,{\smcap i} doublet is the
most powerful for foreground dwarf rejection.  Based on the (admittedly
arbitrary) color and EW(Na) thresholds in Figure~\ref{fig:na_VI0} (dashed
lines), there are 10 and 19~stars in the top right and bottom right sections
which should be considered definite Galactic dwarfs and definite M31 RGB
stars, respectively.  All 29~stars were required to be well measured,
$\rm\sigma[EW(Na)]\lesssim1\AA$, and to be separated by at least
$\rm1.5\sigma[EW(Na)]$ from the horizontal line.  The Na\,{\smcap i} doublet
feature is easily visible for the 10~definite dwarfs in
Figure~\ref{fig:spec_indiv} (marked by triangles).  The definite dwarfs and
definite RGB stars are marked as crosses and filled circles, respectively, in
Figure~\ref{fig:vel_gprob}.  It is very reassuring that these two~subgroups
maintain perfect separation in radial velocity---in other words, none of the
candidate Galactic dwarfs turned out to be a definite M31 RGB star and vice
versa.

It is worth checking whether {\it any\/} of the 16~candidate foreground
Galactic dwarfs might actually be M31 RGB stars.  With this in mind, we
examine their properties using Figures~\ref{fig:cmd}, \ref{fig:vel_gprob},
and~\ref{fig:na_VI0}, going through the subsample one star at a time trying
to identify possible RGB interlopers:
\begin{itemize}
\item[$\bullet$]{Obviously, the 10~cool stars that lie well above the EW(Na)
threshold are definite dwarfs.  It is useful to check how they fare with
respect to some of the other diagnostics.  All 10 lie at or above M31's RGB
tip in the CMD.  Three of them have $P_{\rm giant}\gtrsim0.7$, two have
$P_{\rm giant}\approx0.4$, and five have $P_{\rm giant}<0.1$, again
emphasizing that the $P_{\rm giant}$ parameter is not a perfect discriminator
between dwarfs and RGB stars.}
\item[$\bullet$]{The candidate cool dwarf at $(V-I)_0=1.98$ appears to have a
strong Na\,{\smcap i} doublet feature, $\rm{EW(Na)=3.19\AA}$, but its
spectrum is so noisy that its offset above the threshold corresponds to only
$\rm1\sigma[EW(Na)]$.  In other words, the Na test suggests that it is a
dwarf but is not definitive.  The object lies well below M31's RGB tip in the
CMD.  Its relatively low $P_{\rm giant}$ value of~0.35 tips the scale in
favor of a Galactic dwarf classification.}
\item[$\bullet$]{Two~candidate dwarfs lie close to the EW(Na) threshold, one
0.5~mag redder than the color cut and the other 0.2~mag bluer than it.  Even
though the Na diagnostic is inconclusive, they are both likely to be dwarfs
based on two~other factors: (i)~relatively small $P_{\rm giant}$
values---0.41 and 0.09, respectively; and (ii)~CMD locations close to M31's
RGB tip.}
\item[$\bullet$]{The candidate dwarf with $(V-I)_0=1.46$ and
$\rm{EW(Na)=0.65\AA}$ is too hot for the Na test to be used.  It is about
1~mag fainter than M31's RGB tip and has a relatively high $P_{\rm giant}$
value, 0.63.  These factors suggest that the object may be an M31 RGB star.}
\item[$\bullet$]{Finally, two~candidate dwarfs lie within $\approx0.1$~mag of
the color cut but on the blue side.  Both have noisy spectra but appear to
lie signficantly below the threshold if our empirical scaling of
$\rm\sigma[EW(Na)]$ is to be trusted.  It is intriguing that the two have
relative large $P_{\rm giant}$ values, 0.54 and 0.86, and both lie
$\approx1$~mag below the tip of M31's RGB in the CMD.  Like the last star,
these two should also be considered possible M31 RGB stars.}
\end{itemize}
In summary, there are three~stars among the 16~velocity-selected candidate
dwarfs that could be M31 RGB stars.  All three are relatively faint and do
not have well measured photometric and spectroscopic parameters as a result.
None of the three present a strong/definite enough case to warrant
reclassification at this stage.  We will return to a discussion of these
objects in \S\,\ref{sec:halo_dyn}.

We next use similar reasoning to test whether {\it any\/} of the 68~candidate
M31 RGB stars might actually be foreground Galactic dwarfs:
\begin{itemize}
\item[$\bullet$]{The 48~candidate M31 RGB stars with $v<-400$~km~s$^{-1}$ can
be ruled out as potential dwarfs on the basis of the IASG Galactic structure
model \citep{bah84,rat85} which predicts that the foreground dwarf
contamination rate should be negligible at these large negative radial
velocities (see Fig.~5 of RG02).  The giant southern stream population should
be completely free of foreground dwarf contamination by virtue of its large
blueshift.}
\item[$\bullet$]{The remaining 20~candidate M31 RGB stars in the radial
velocity range $-400<v<-200$~km~s$^{-1}$ all have $P_{\rm giant}\gtrsim0.6$
so are unlikely to be dwarfs.}
\item[$\bullet$]{The Na diagnostic does not turn up any compelling evidence
to suggest that there dwarf interlopers among this sample of 20.  Twelve are
to the left of the color cut and eight are to the right of it.  None of the
20 are significantly above the EW(Na) threshold; six are significantly below
it, three on each side of the color cut.}
\end{itemize}
In summary, there is nothing to suggest that any of the candidate M31 RGB
stars requires reclassification.

Given the clean RGB versus dwarf star distinction provided by the Na\,{\smcap
i} doublet diagnostic in Figures~\ref{fig:vel_gprob} and \ref{fig:na_VI0},
coupled with all the earlier evidence, it is probably safe to drop the term
``candidate'' from the designations: Unless otherwise mentioned, we refer to
the $v<-200$~km~s$^{-1}$ group as M31 RGB stars and the $v>-150$~km~s$^{-1}$
group as Galactic dwarfs throughout the rest of this paper.

\subsubsection{Comparison to Earlier Studies}\label{sec:reject_previous}

Previous spectroscopic studies of M31 stars have adopted a more limited
approach to RGB versus dwarf star discrimination.  None of these studies had
access to $DDO51$ photometry that is so valuable for suppressing foreground
dwarf contamination in our DEIMOS survey.  RG02 used radial velocities, CMD
information, and the $\rm[Fe/H]_{spec}$ versus $\rm[Fe/H]_{phot}$ comparison;
the resulting RGB/dwarf star separation was not as clear as in this study.
The \citet{rei04} study used radial velocities alone, while \citet{iba04}
made no mention of foreground contamination; neither of these last
two~studies made use of spectral absorption line strengths.

\section{Dynamics}\label{sec:dyn}

The dynamics of the giant southern stream and general halo (non-stream)
populations in M31 are described in this section.  A maximum likelihood fit
of the sum of two~Gaussians is carried out on the radial velocity
distribution of the 68~M31 RGB stars.  This is used to characterize the mean
velocity and velocity dispersion of the two~components.
Figure~\ref{fig:maxlik} shows $\Delta\chi^2$ ($\equiv\chi^2-\chi_min^2$)
curves for these four~quantities (four of the five~parameters in the fit);
the optimal value of each parameter and uncertainty, in the form of 90\%
confidence limits, are indicated.

\subsection{Giant Southern Stream}\label{sec:strm_dyn}

\subsubsection{Mean Radial Velocity}\label{sec:strm_mean_vel}

The taller/narrower of the two~Gaussians in the maximum likelihood fit
corresponds to the main peak in the stellar radial velocity distribution
[Fig.~\ref{fig:vhist}({\it d\/})].  Based on this fit, we determine that the
giant southern stream has a mean heliocentric radial velocity of
$-458\pm6$~km~s$^{-1}$ (90\% confidence limits), or a line-of-sight velocity
of $-158$~km~s$^{-1}$ with respect to M31 [see Fig.~\ref{fig:maxlik}({\it
a\/})].  Our measurement is consistent with recent measurements by
\citet{lew04} and \citet{iba04} at other points along the stream in the sense
that our field~`a3' lies between their fields~2 and~6 both in terms of sky
position and radial velocity.  The companion Paper~II uses the field~`a3'
mean velocity from this study, along with other velocity data
\citep{lew04,iba04} and line-of-sight distance estimates \citep{mcc03}, to
constrain the orbit of the stream and its progenitor satellite.

\subsubsection{Velocity Gradients}\label{sec:vel_grad}

\begin{figure}
\centerline{\epsfxsize=7in \epsfysize=9in
\epsfbox{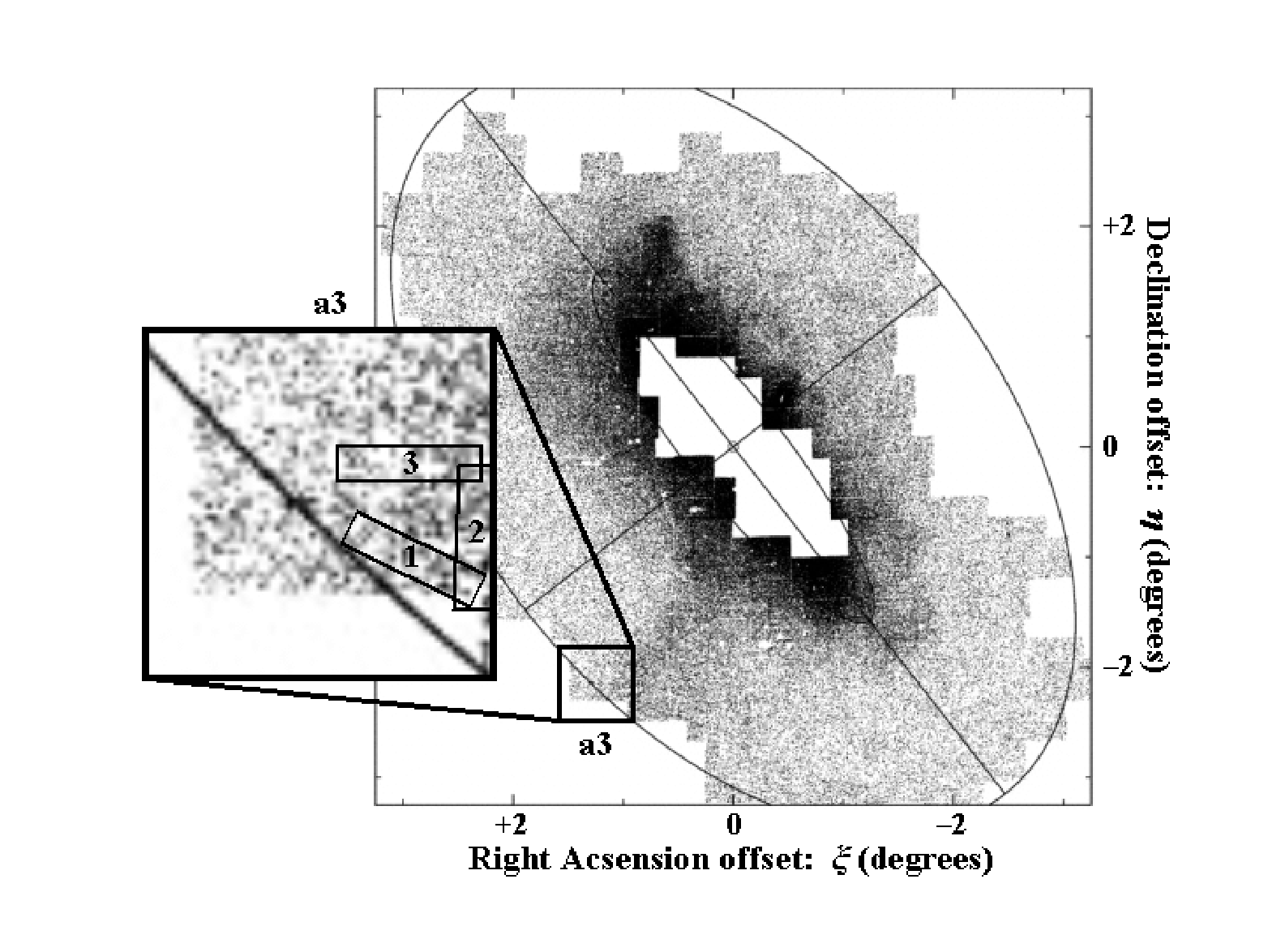}}
\caption{\label{fig:fld_loc}{
Sky position and orientation of the DEIMOS spectroscopic slitmasks (thin
rectangles numbered 1--3 in the inset), against the star-count map of
metal-rich M31 RGB stars presented by \citet{fer02}.  A DEIMOS slitmask
covers an area of roughly $16'\times4'$.  The bold square (main plot and
inset) indicates the full $35'\times35'$ area of field~`a3' from the KPNO
4-m/MOSAIC imaging survey of \citet{ost02}; our spectroscopic targets were
selected from the $DDO51$-, $M$-, and $T_2$-band photometric catalog of
objects in this field.  The background ellipse (also visible in the inset)
indicates the shape of M31's stellar halo and marks the rough extent of the
published star-count map.  The sharp northeast edge of the giant southern
stream in M31 runs more or less diagonally through the middle of field~`a3'
so that the southwest half of the field is on the stream; the three~DEIMOS
masks lie mostly on the stream.
}}
\end{figure}

\begin{figure}
\centerline{\epsfxsize=5.2in \epsfysize=5.2in
\epsfbox{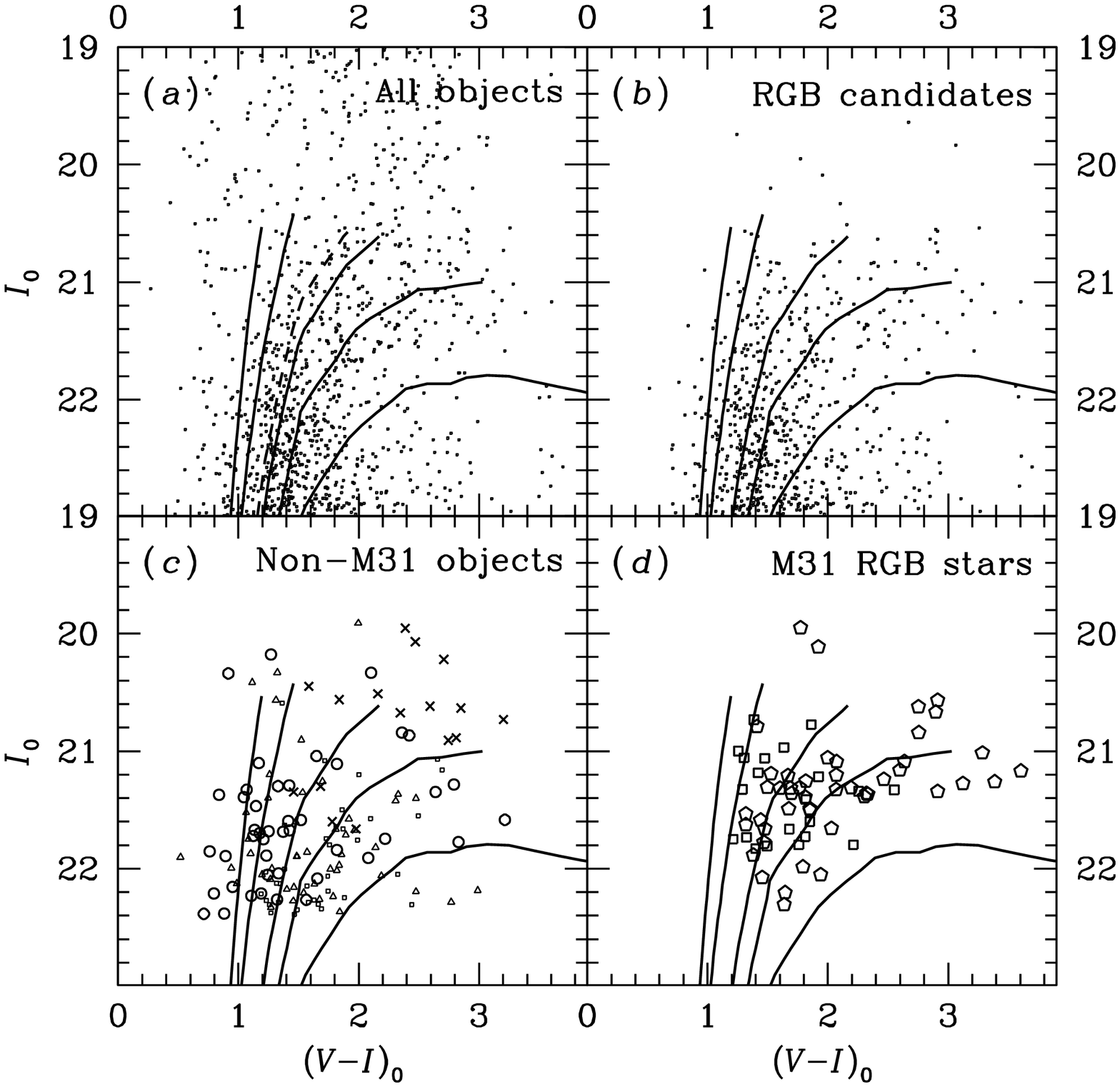}}
\caption{\label{fig:cmd}{
({\it a\/})~Color-magnitude diagram of all objects in the portion of the `a3'
field covered by the three~DEIMOS slitmasks.  The Washington system $M$ and
$T_2$ photometry has been transformed to Johnson/Cousins $V$ and $I$
\citep{maj00} and corrected for foreground extinction/reddening
\citep{sch98}.  Red giant branch fiducials from the Padova group
\citep{gir00} are shown for $t=12.6$~Gyr and $\rm[Fe/H]=-2.3$, $-1.3$,
$-0.7$, $-0.4$, and 0.0 (solid lines, left$\rightarrow$right) and $t=6.3$~Gyr
and $\rm[Fe/H]=-0.7$ (dashed line).~~
({\it b\/})~Same as ({\it a\/}), except the $DDO51$-based parameter $P_{\rm
giant}>0.5$ is used to pre-select RGB candidates and the morphological
selection criteria ${\tt chi}<1.3$ and $\vert{\tt sharp}\vert<0.3$ are used
to reject background galaxies (\S\,\ref{sec:targ_sel}).~~
({\it c\/})~Same as ({\it a\/}), for objects for which the radial velocity
measurement fails because of low~S/N (small squares) or lack of definite
spectral features (small triangles), foreground Galactic dwarf stars
(crosses), and background field galaxies (open circles); see
\S\,\ref{sec:success} for details.~~
({\it d\/})~Same as ({\it a\/}), for confirmed M31 RGB stars: potential
members of the giant southern stream (open pentagons) and members of the
general halo population (open squares).  The M31 RGB stars are bracketed by
model RGB tracks with a plausible range of metallicities, with the stream RGB
stars being more metal rich on average than those in the general halo.  Nine
stars, all potential stream members, lie at or above the tip of the RGB
tracks.
}}
\end{figure}

\begin{figure}
\centerline{\epsfxsize=6in \epsfysize=6in
\epsfbox{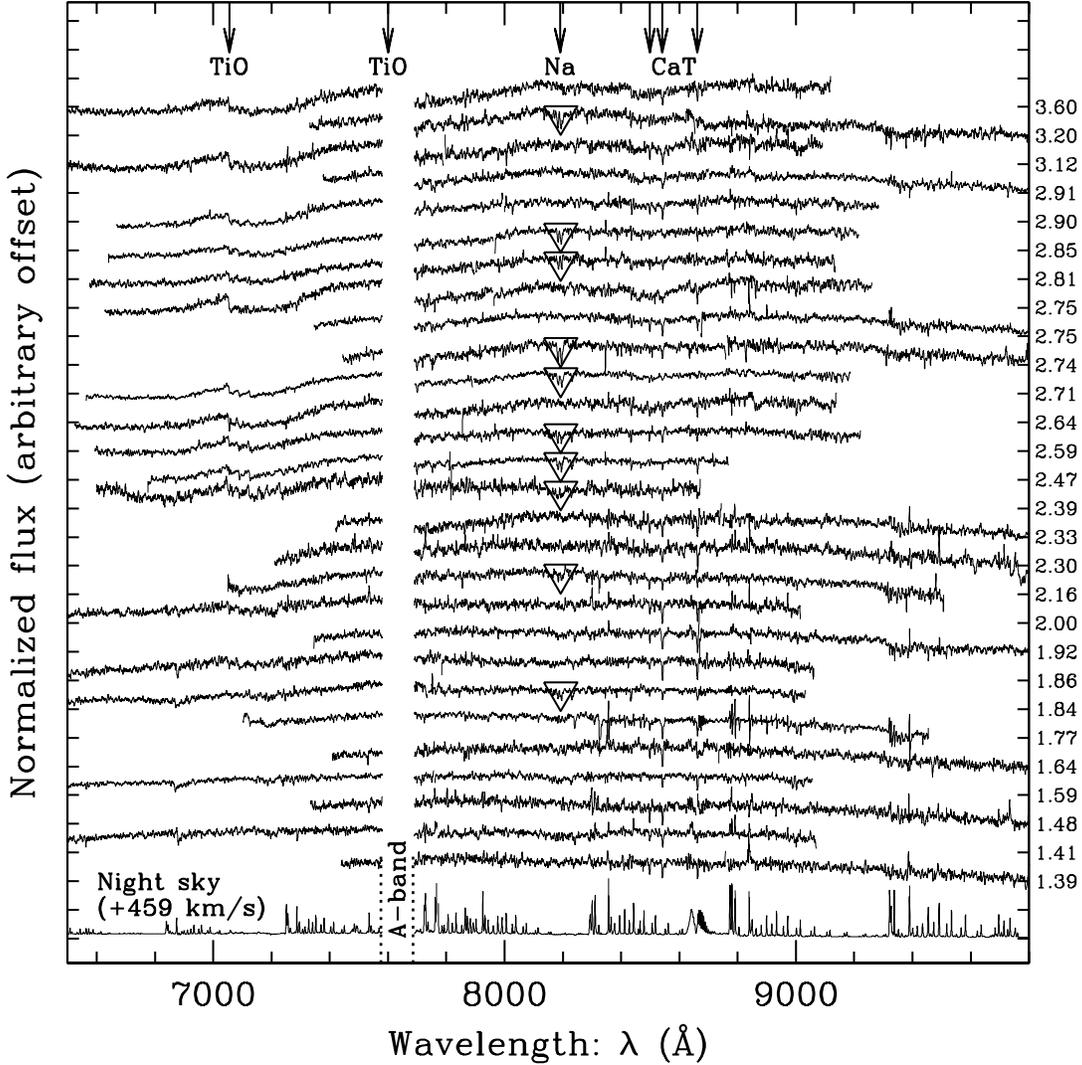}}
\caption{\label{fig:spec_indiv}{
Montage of the best DEIMOS spectra ($\rm S/N\gtrsim12\,\AA^{-1}$) of M31 RGB
and Galactic dwarf stars.  The spectra have been smoothed with a 3\AA\
weighted boxcar, shifted to zero velocity, normalized at
$\rm\sim\lambda8500\AA$, offset in $y$ for illustration purposes (each
tickmark represents unity), and ordered by $(V-I)_0$ color (indicated on the
right side of the $y$ axis).  One of two~instrument settings was used
covering $\lambda\lambda6500$--9100\AA\ or $\lambda\lambda7200$--9800\AA.
The atmospheric A-band correction is inadequate in the present reduction so
this region is excluded.  The night sky spectrum (bottom), a composite from
the two~wavelength settings, has been smoothed and shifted by
$+458$~km~s$^{-1}$, the mean shift for stars in M31's giant southern stream.
The strongest TiO bandheads, Na\,{\smcap i} doublet, and Ca\,{\smcap ii}
triplet lines are marked.  The TiO bands increase in strength with $(V-I)_0$
color; the redder TiO band lies in the A-band gap but the break in the
spectrum is apparent for red stars.  The Na line is surface-gravity sensitive
and thus a discriminator between M31 RGB stars and foreground Galactic dwarfs
for $(V-I)_0\gtrsim2$ (latter marked by bold open triangles).
}}
\end{figure}

\begin{figure}
\centerline{\epsfxsize=6.2in \epsfysize=6.2in
\epsfbox{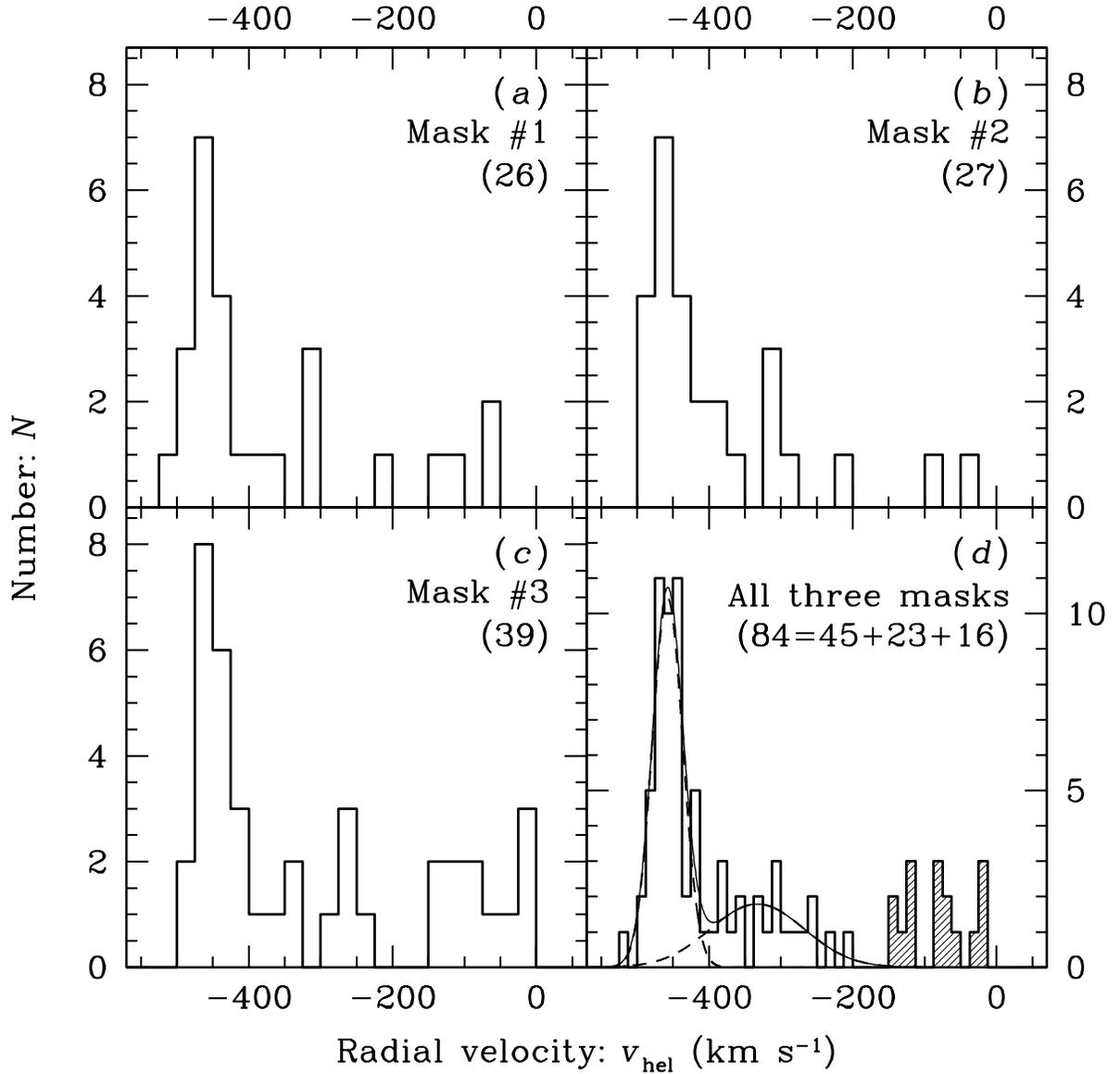}}
\caption{\label{fig:vhist}{
({\it a\/})--({\it c\/})~Distribution of (heliocentric) radial velocities for
stars on each of the three~DEIMOS masks.  The total number of objects in each
histogram is indicated in parentheses.~~
({\it d\/})~Same as ({\it a\/}), for all three~masks combined (note, 8~stars
have duplicate measurements).  The thin dashed lines show Gaussians
representing M31 RGB stars in the giant southern stream (tall narrow peak)
and general halo (short broad peak); the thin solid line is the sum of the
two~Gaussians (see Fig.~\ref{fig:maxlik} for the details of the maximum
likelihood fit).  The group of objects with $v>-150$~km~s$^{-1}$ are
candidate foreground Galactic dwarf stars (shaded histogram).  The likely
break up of the 84~stars is: 45~M31 stream RGB stars, 23~M31 halo RGB stars,
and 16~foreground Galactic dwarfs
(\S\S\,\ref{sec:contrast}--\ref{sec:reject}).
}}
\end{figure}

\begin{figure}
\centerline{\epsfxsize=7in \epsfysize=7in
\epsfbox{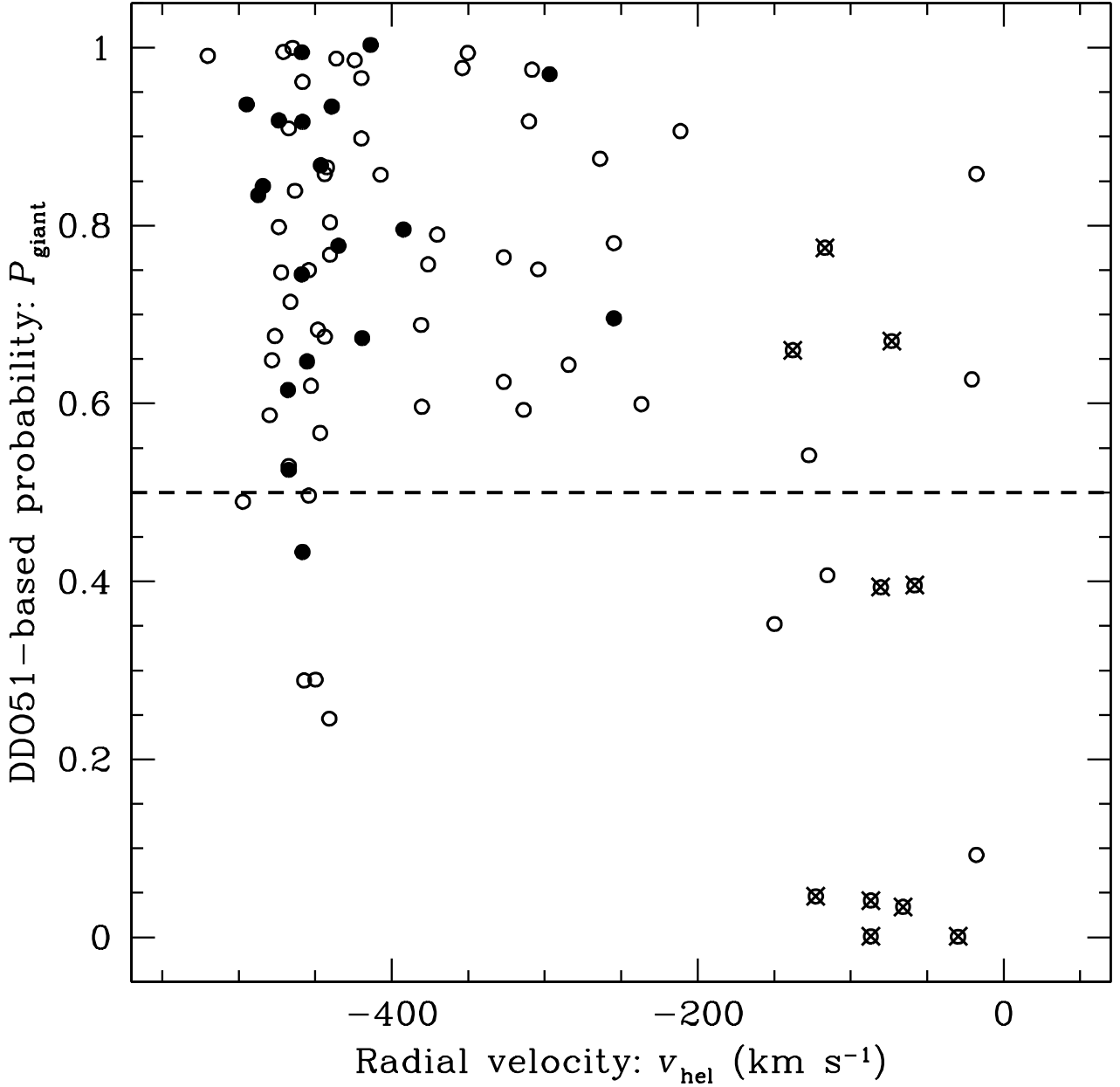}}
\caption{\label{fig:vel_gprob}{
$DDO51$-based probability of a star being a red giant plotted against radial
velocity.  Stars with $v>-150$~km~s$^{-1}$ are foreground Galactic dwarfs
while the rest are M31 RGB stars.  The dashed line separates
primary/secondary spectroscopic targets with $P_{\rm giant}>0.5$ (lists~1
and~2) from ``filler'' targets with $P_{\rm giant}<0.5$ (lists~3 and~4; see
\S\,\ref{sec:targ_sel}).  For a subset of the stars (red stars with
high~S/N), the surface-gravity-sensitive $\rm\lambda8190\AA$ Na\,{\smcap i}
doublet is used to make a definite classification of dwarfs (crosses) versus
RGB stars (filled circles); see \S\,\ref{sec:reject_clean} and
Figure~\ref{fig:na_VI0} for details.  The $P_{\rm giant}$ values tend to be
higher for confirmed RGB stars than dwarfs but it is not a perfect
discriminator.
}}
\end{figure}

\begin{figure}
\centerline{\epsfxsize=6.2in \epsfysize=6.2in
\epsfbox{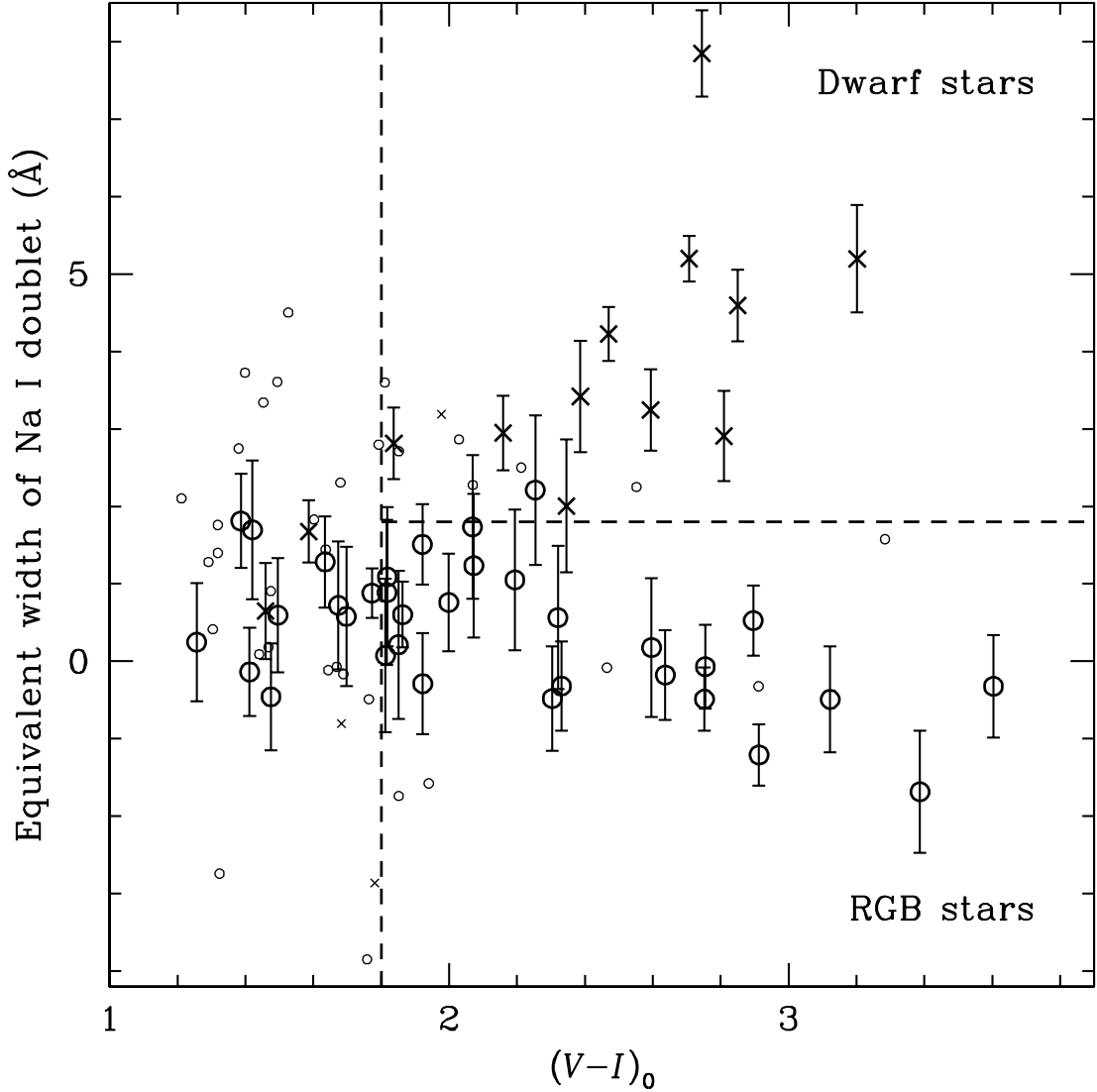}}
\caption{\label{fig:na_VI0}{
Equivalent width of the Na\,{\smcap i} doublet versus dereddened color for
the 84~stars in our sample.  The EW is calculated within a 21\AA-wide window.
The uncertainty in line strength is estimated from stars with duplicate
measurements under the assumption that it is inversely proportional to S/N;
$1\sigma$ error bars are only shown when they are $\rm<1\AA$.  The
velocity-based subsamples of candidate M31 RGB stars and candidate Galactic
dwarf stars are marked as crosses and circles, respectively.  Cool stars
[$T_{\rm eff}<4000\,$K or $(V-I)_0>1.8$] have a bimodal distribution of
Na\,{\smcap i} line strengths.  The dashed lines are used to discriminate
between RGB and dwarf stars---they are expected to occupy the bottom right
and top right sections of the plot, respectively.
}}
\end{figure}

\begin{figure}
\centerline{\epsfxsize=6.2in \epsfysize=6.2in
\epsfbox{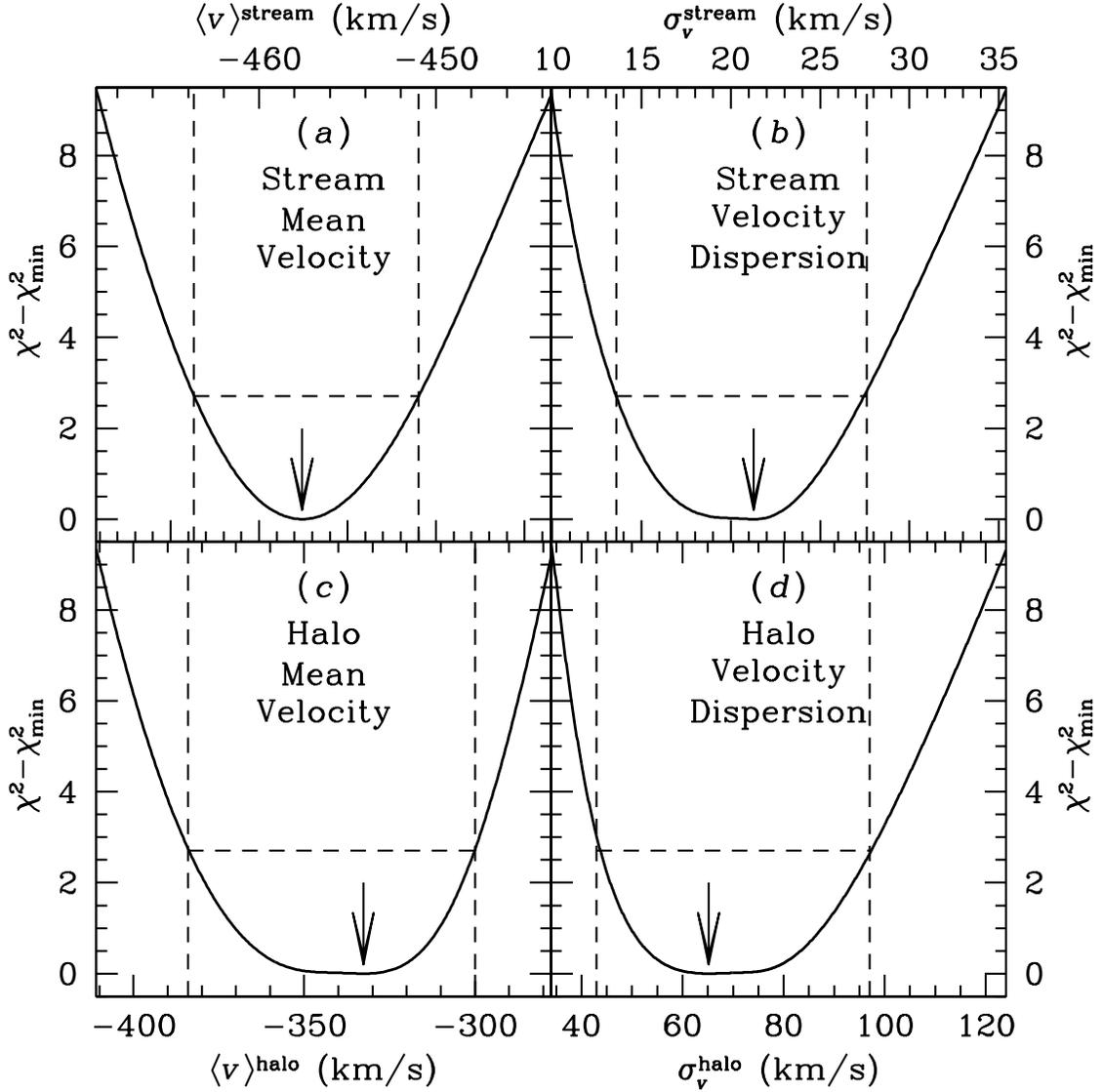}}
\caption{\label{fig:maxlik}{
Results of a maximum likelihood fit of the sum of two~Gaussians to the radial
velocity distribution of M31 RGB stars.  The difference between $\chi^2$ and
the minimum $\chi^2$ is plotted as a function of ({\it a\/})~the mean
velocity of the giant southern stream, ({\it b\/})~the stream velocity
dispersion, ({\it c\/})~the mean velocity of the general halo population, and
({\it d\/})~the halo velocity dispersion.  An arrow marks the best fit value
and the dashed lines corresponds to the 90\% confidence limits in each case.
}}
\end{figure}

\begin{figure}
\centerline{\epsfxsize=6.2in \epsfysize=6.2in
\epsfbox{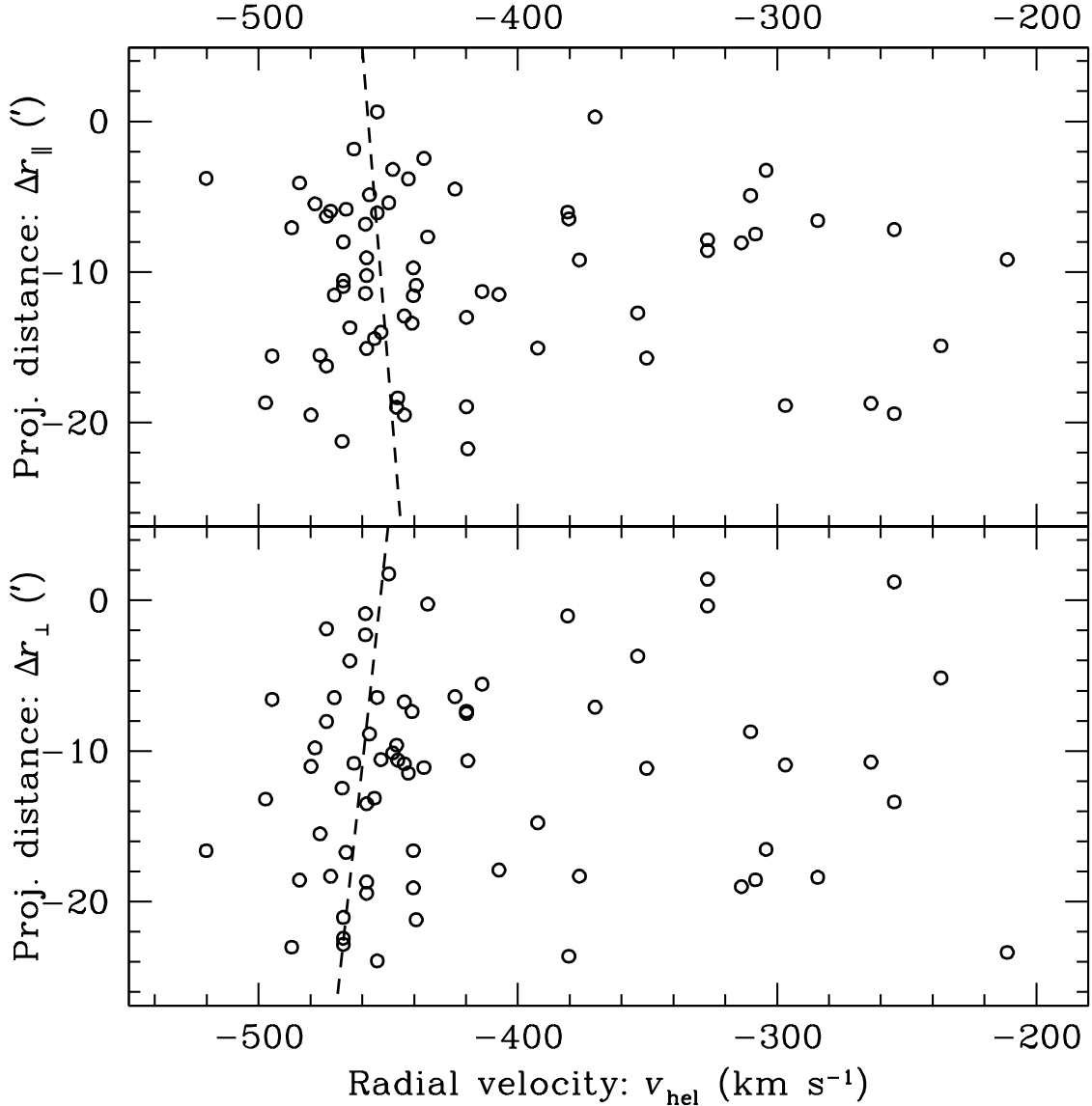}}
\caption{\label{fig:vgrad}{
Projected position of M31 RGB stars along ({\it top\/}) and perpendicular to
({\it bottom\/}) the main length of the giant southern stream plotted against
radial velocity.  The stream is assumed to run northwest to southeast, and
$\Delta{r}_\parallel$ and $\Delta{r}_\perp$ are measured relative to the
center of the `a3' field increasing towards the southeast and northeast,
respectively (Fig.~\ref{fig:fld_loc}).  The dashed lines indicate the
gradients $\Delta{v}/\Delta{r}$ measured for candidate stream stars
($v<-410$~km~s$^{-1}$); the gradients are poorly determined because of
uncertainty about which stars are members of the stream (as opposed to the
general M31 halo) and because of the small number of stars.  Our
$\Delta{v}/\Delta{r}_\parallel$ measurement is consistent with that measured
over a longer spatial baseline by \citet{lew04} and \citet{iba04}; large
values of $\vert\Delta{v}/\Delta{r}_\perp\vert$ are ruled out by the data.
}}
\end{figure}

We next explore possible trends in the mean velocity of M31's giant southern
stream as a function of sky position.  Figure~\ref{fig:vgrad} shows the
radial velocity distribution of M31 RGB stars as a function of projected
position along and perpendicular to the stream (top and bottom,
respectively).  The positions are defined relative to the center of
field~`a3' with $\Delta{r}_\parallel$ increasing from northwest to southeast
and $\Delta{r}_\perp$ increasing from southwest to northeast; in other words,
the projection of the stream is assumed to be at a position angle of
$-45^\circ$ (Fig.~\ref{fig:fld_loc}).  We conclude that there is no strong
velocity gradient along either axis.  The dashed lines in
Figure~\ref{fig:vgrad} mark the nominal slopes,
$dv/dr_\parallel=-0.5$~km~s$^{-1}$~arcmin$^{-1}$ and
$dv/dr_\perp=+0.6$~km~s$^{-1}$~arcmin$^{-1}$, but there is uncertainty about
the membership of any given RGB star (stream versus general halo; see
\S\,\ref{sec:contrast}) and large Poisson errors.  As discussed in Paper~II,
our measurement of the local velocity gradient along the stream in field~`a3'
is consistent with the published radial velocities over a longer spatial
baseline \citep{lew04,iba04}.

Any systematic trend in the stream's mean velocity with position tends to
broaden the peak in the velocity histogram.  The radial velocity gradient
along the direction parallel to the stream is
$dv/dr_\parallel=-0.47$~km~s$^{-1}$~arcmin$^{-1}$ so this translates to a
spread of $\Delta{v}=\pm4.7$~km~s$^{-1}$ over the
$\Delta{r}_\parallel\sim20'$ spanned by our three~masks along the length of
the giant southern stream (see Figs.~\ref{fig:fld_loc} and~\ref{fig:vgrad}).
This spread is small compared to the width of the peak associated with the
stream.  Indeed we have checked explicitly that correcting for the gradient
has a negligible effect on the width of the best-fit Gaussian.

\subsubsection{Intrinsic Velocity Dispersion}\label{sec:vel_disp}

Figure~\ref{fig:maxlik}({\it b\/}) shows the likelihood function for the
width of the Gaussian corresponding to the giant southern stream.  The
best-fit width is $\sigma_v^{\rm stream}=21\pm7$~km~s$^{-1}$ (90\% confidence
limits).  The $1\sigma$ velocity measurement error of 15~km~s$^{-1}$
(\S\,\ref{sec:vel_error}) is subtracted in quadrature from the measured
velocity dispersion of the stream.  Our best estimate of the {\it
intrinsic\/} line-of-sight velocity dispersion of the stream is
$\sigma_v^{\rm stream}({\rm intrinsic})\approx15$~km~s$^{-1}$ but, given the
large uncertainty in the measured value, we conclude that $\sigma_v^{\rm
stream}({\rm intrinsic})\lesssim23$~km~s$^{-1}$ (90\% confidence limit).
This is comparable to the velocity dispersions measured in the Milky Way's
Monoceros and Sagittarius streams \citep{cra03,maj04a}.  We consider the
coldness of the stream and its implications in Paper~II.

\subsection{General Halo (Non-Stream) Population: Evidence of
Substructure}\label{sec:halo_dyn}

In this section we turn our attention to the kinematics of M31's general halo
(i.e.,~non-stream) population.  The broad, low-level component in the
combined stellar radial velocity histogram [Fig.~\ref{fig:vhist}({\it d\/})]
spans the range $-500\lesssim{v}<-200$~km~s$^{-1}$ and extends to the right
of the main peak.  As we will show in \S\,\ref{sec:feh_comp} below this
component has a large spread in [Fe/H] comparable to that seen in other
studies of the halo (even though there are differences in detail).  The broad
component is {\it not\/} merely the tail of the giant southern stream's
radial velocity distribution judging from the differences in their stellar
populations: The former appears to be more metal poor, has a larger
metallicity spread, and lacks stars above the RGB tip (see
\S\,\ref{sec:feh_comp} and Figs.~\ref{fig:cmd} and \ref{fig:feh}).

The broader of the two~components in the maximum likelihood fit to the radial
velocity distribution of M31 RGB stars is centered at a heliocentric velocity
of $-333_{-51}^{+33}$~km~s$^{-1}$ and has a Gaussian width of
$65_{-21}^{+32}$~km~s$^{-1}$ [90\% confidence limits; see
Fig.~\ref{fig:maxlik}({\it c\/}--{\it d\/})].  The broad halo component
appears to be adequately fit by a Gaussian but the large Poisson fluctuations
(only $\approx21$~stars belong to this component) make it impossible to tell
whether its radial velocity distribution is drawn from a truly smooth and
virialized underlying distribution or if it contains substantial
substructure.

The velocity distribution of the general halo population in field~`a3'
appears to be anomalously narrow.  Its width $\sigma_v^{\rm halo}({\rm
field}~$`a3'$)=65_{-21}^{+32}$~km~s$^{-1}$ is significantly smaller than the
width of $\sigma_v^{\rm halo}({\rm other})=150$~km~s$^{-1}$ measured for
other M31 halo tracers: field RGB stars in an $R=19$~kpc minor-axis field
(RG02) and global samples of globular clusters and planetary nebulae
\citep[see][and references therein]{eva00}.  A plausible explanation of this
is that our field~`a3' sample is dominated by substructure in the
halo--e.g.,~it consists of one or two subclumps instead of a smooth,
virialized distribution.  This is the first of three~lines of evidence
pointing to the possible existence of substructure in the general halo
population in this remote M31 field.

We next consider the three~possible M31 RGB interlopers among the 16~stars
with $v>-150$~km~s$^{-1}$ (candidate Galactic dwarfs) that have been excluded
from the dynamical analysis so far (see bullets in
\S\,\ref{sec:reject_clean}).  Their radial velocities, $-21$, $-127$, and
$-18$~km~s$^{-1}$, place them right among the foreground Galactic dwarf
population, but it is possible that they belong to the M31 halo instead: For
example, they would be within $3\sigma_v^{\rm halo}$ of a broad distribution
centered on M31's systemic velocity of $-300$~km~s$^{-1}$.  Including all
three as members of the M31 halo changes the best-fit width to $\sigma_v^{\rm
halo}=116_{-22}^{+31}$~km~s$^{-1}$, while including only the last pair of
stars as halo members (probably more realistic) changes the width to
$\sigma_v^{\rm halo}=103_{-21}^{+28}$~km~s$^{-1}$, where the quoted
uncertainties are 90\% confidence limits derived from new maximum likelihood
fits in each case.  Our basic conclusion about the field~`a3' halo sample
having an unusually narrow velocity distribution seems to be independent of
whether or not the possible RGB interlopers are included in the sample.

The centroid of the radial velocity distribution of the halo population is
offset by $-33$~km~s$^{-1}$ with respect to M31's systemic velocity of
$-300$~km~s$^{-1}$, and the offset is significant at the 90\% confidence
level, based on the maximum likelihood fit to the primary sample of 68~M31
RGB stars (possible interlopers excluded).  Inclusion of the possible
interlopers leads to a small shift in the best-fit halo mean velocity towards
less negative values and a reduction therefore in the magnitude of the offset
relative to systemic, along with a slight increase in the uncertainty: The
net result of these changes is the best-fit offset of $-19$~km~s$^{-1}$ is no
longer statistically significant.  If the observed offset is
real/significant, it could be a second sign of substructure in M31's halo.  A
third possible sign of substructure is the apparent difference between the
metallicity distributions of the field~`a3' halo sample and RG02's sample
(see \S\,\ref{sec:feh_comp} below).

An alternative explanation of the observed velocity offset (taken at face
value) is that it is a result of global rotation of M31's halo.  The sense of
halo rotation would then be the same as for M31's disk whose southwest half
is known to be blueshifted relative to systemic.  Since our field~`a3' lies
close to the minor axis, this would imply that we are seeing only a small
component of the full halo rotation speed and would therefore place a lower
limit of $(v_{\rm rot}/\sigma_v)^{\rm halo}\gtrsim0.2$.  The M31 halo appears
to have an aspect ratio of 5:3 (in projection) judging from the isopleths in
the \citet{iba01} star-count map.  If the observed flattening is due to halo
rotation, this would require $(v_{\rm rot}/\sigma_v)^{\rm halo}\sim0.8$
\citep{bin87} which is consistent with our lower limit.  If M31's halo is
confirmed to be rotating at this level, it would be in stark contrast to the
Milky Way halo which appears to have little or no (or even slight retrograde)
rotation \citep*{maj92,maj96,pop98}.

In closing, we note that there are very few constraints on the global
dynamics of M31's stellar halo at the present time.  Only two~sight-lines
have been probed using field RGB stars---field~`a3' from this study, at a
projected distance of $R=31$~kpc slightly off the minor axis, and RG02's 
19~kpc minor-axis field---and both studies are hampered by small number
statistics.  Future papers from our survey will use all of the existing
fields with DEIMOS spectra (about half a dozen locations scattered around
M31) and it should be possible to derive tighter constraints on the dynamics
of the halo.

\section{Chemical Abundance (and Age)}\label{sec:feh}

The giant southern stream in M31 appears to be the remnant of a
massive/luminous satellite.  The chemical abundance and age distribution of
the stream are best viewed in the context of the ensemble of former
satellites that merged to form the galaxy's halo.  With this goal in mind, we
compare the properties of luminous RGB stars in the stream and to those in
the general halo.  The analysis of RGB star metallicities in this section is
based mostly on their photometric properties with limited use of
spectroscopic information; a full analysis of spectroscopic chemical
abundances will be presented in a future paper.  Some indirect age
constraints are also obtained in this section.

\subsection{Photometric Metallicity Estimates}\label{sec:feh_phot}

\subsubsection{Measurement Method and Errors}\label{sec:feh_phot_method}

Photometric metallicity estimates are obtained for the 68~confirmed M31 RGB
stars in our field~`a3' sample.  As illustrated in Figure~\ref{fig:cmd}({\it
d\/}), the position of each star in the $I_0$ versus $(V-I)_0$ CMD is
compared to a set of model RGB fiducials \citep{gir00}.  The conversion of
stellar photometry from the Washington system to the Johnson/Cousins system
and star-by-star dereddening are described in \S\,\ref{sec:targ_sel}.  The
fiducials span a wide range of metallicities for an age of $t=12.6$~Gyr; they
have been placed on the CMD using a true distance modulus of $(m-M)_0=24.47$
corresponding to an adopted distance to M31 of~783~kpc \citep{sta98,hol98}.
In order to estimate $\rm[Fe/H]_{phot}$ for the stars, a Legendre polynomial
of 6th order in $(V-I)_0$ and 10th order in $I_0$ is used to interpolate
between the model RGB tracks.

If an M31 star is~2--6~Gyr old instead of our assumed age of~12.6~Gyr, its
photometric metallicity estimate would need to be revised upwards by about
$+0.3$~dex [e.g.,~a $t=6.3$~Gyr fiducial is shown as a dashed line in
Fig.~\ref{fig:cmd}({\it a\/})].  The overall uncertainty in the metallicities
is about~0.3~dex, dominated by systematic errors such as age
error/spread, residual differential reddening, [$\alpha$/Fe] variations, and
model inaccuracies; however, relative metallicity rankings can be achieved to
somewhat greater accuracy than this for stars of comparable age.  For stars
located above the RGB tip in the CMD, the $\rm[Fe/H]_{phot}$ estimates are
based on linear extrapolation and are therefore very uncertain.

\subsubsection{Selection Biases}\label{sec:feh_bias}

Before studying the metallicity distribution of M31 stars, we investigate
whether our RGB star sample is an unbiased sample.  A couple of selection
effects in particular are worth discussing.

It is well known that metal-line blanketing causes the RGB tip gets fainter
with increasing metallicity.  This causes the most metal-rich of the luminous
RGB stars, those with $\rm[Fe/H]\gtrsim-1$, to be underrepresented in any
magnitude-limited sample.  RG02 characterized this bias in detail and
corrected for it.  We do not correct for the bias in this paper since we are
primarily interested in a differential stream versus general halo comparison
rather than the absolute shape of the [Fe/H] distribution.  Moreover, the
bias should be less pronounced in our field~`a3' sample (and the rest of the
DEIMOS survey) than in RG02's sample: they used a limiting $I$-band magnitude
of~22.0, whereas our study used~22.0 for list~1 primary targets but relaxed
it to~22.5 for lists~2--4 secondary and ``filler'' targets
(\S\,\ref{sec:dsim}).

\begin{figure}
\centerline{\epsfxsize=6.5in \epsfysize=6.5in
\epsfbox{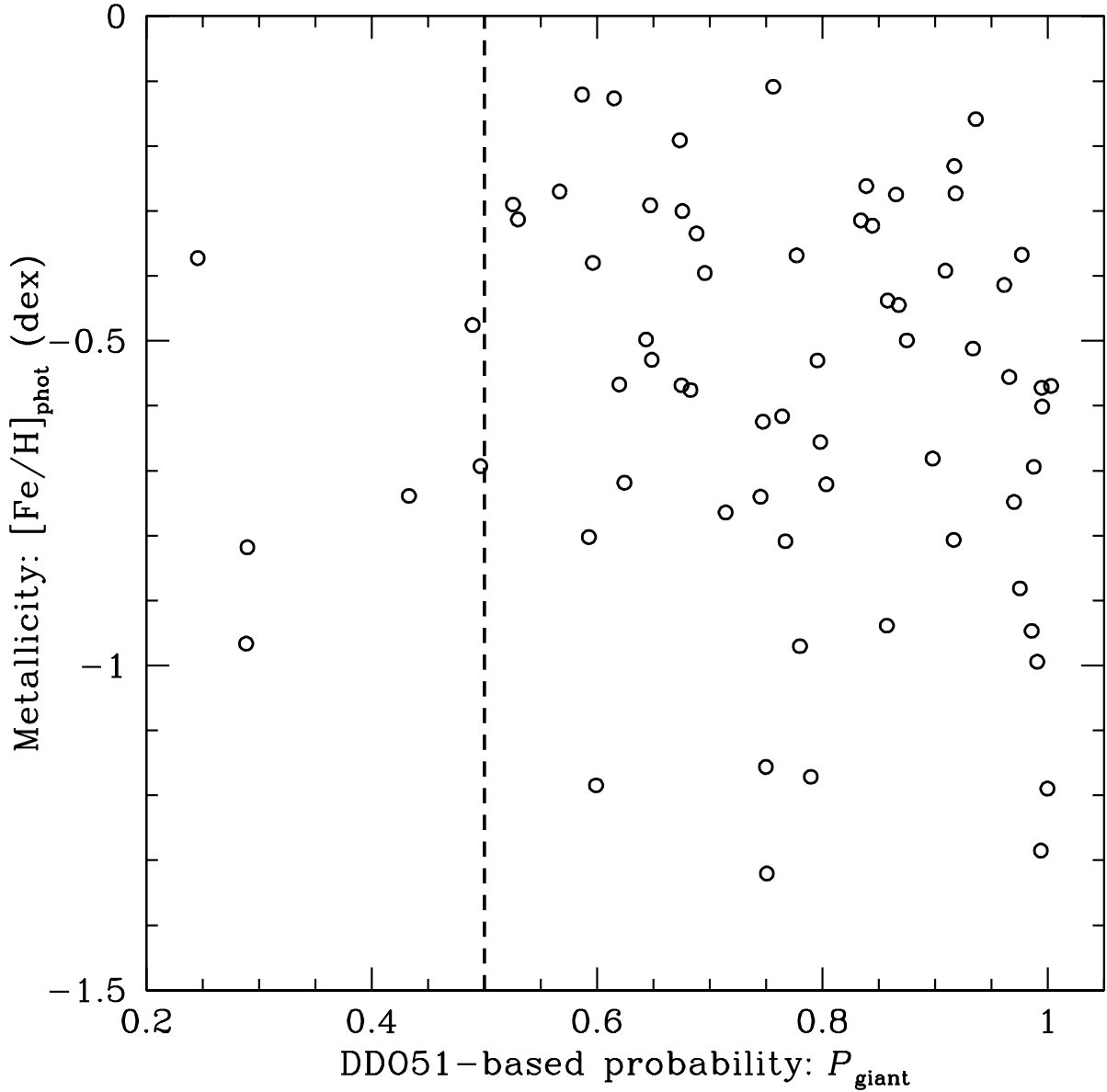}}
\caption{\label{fig:gprob_feh}{
Photometric metallicity estimate versus $P_{\rm giant}$, the $DDO51$-based
probability of a star being a red giant.  Only confirmed M31 RGB stars, those
with $v<-200$~km~s$^{-1}$ are shown.  Stars to the right of the dashed
vertical line are primary and secondary RGB candidates (lists~1 and~2) while
the ones to the left are ``filler'' targets (lists~3 and~4; see
\S\,\ref{sec:dsim}).  Two~features of this plot suggest that $DDO51$-based
pre-screening does {\it not\/} introduce any strong metallicity bias:
(i)~Stars with $P_{\rm giant}>0.5$ occupy a roughly rectangular region of the
plot and show no obvious trend; and (ii)~Even though there is only a handful
of ``filler'' targets---i.e.,~those for which the $DDO51$ criterion was
relaxed or dropped---they span the same range of $\rm[Fe/H]_{\rm phot}$
values as the rest of the stars.
}}
\end{figure}

\begin{figure}
\centerline{\epsfxsize=6.5in \epsfysize=6.5in
\epsfbox{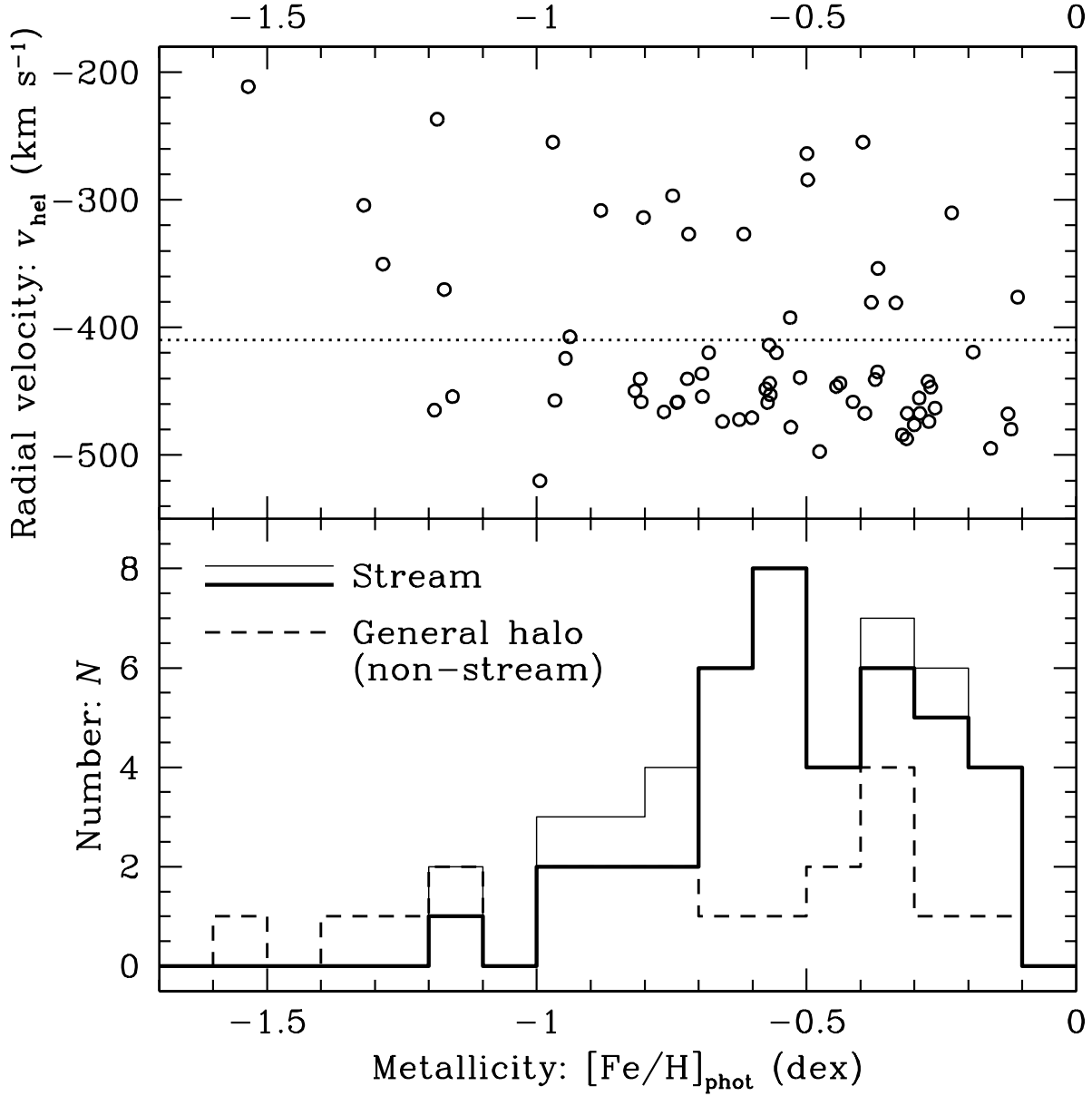}}
\caption{\label{fig:feh}{
({\it Top\/})~Radial velocity versus metallicity for M31 RGB stars, where the
photometric metallicity estimate is derived by fitting model red giant tracks
to the star in the color-magnitude diagram (Fig.~\ref{fig:cmd}).  The
velocity range below the dotted line is dominated by members of the giant
southern stream.~~
({\it Bottom\/})~Metallicity distribution of M31 RGB stars in the stream's
velocity range, with and without the seven~stars brighter than the RGB tip
for which $\rm[Fe/H]_{phot}$ estimates are unreliable (thin and bold solid
histograms, respectively) and the non-stream general halo population (dashed
histogram).  We estimate that two~of the 47~stars in the stream's velocity
range are actually members of the general halo population
(\S\,\ref{sec:contrast}).  The stream appears to have a smaller metallicity
spread than the general halo and a higher mean metallicity:
$\langle\rm[Fe/H]_{\rm phot}(stream)\rangle\sim-0.51$ versus
$\langle\rm[Fe/H]_{\rm phot}(halo)\rangle\lesssim-0.74$
(\S\,\ref{sec:feh_comp}).
}}
\end{figure}

Since metallicity is the second parameter (after surface gravity) in
determining the strength of the Mg\,b/MgH feature, our $DDO51$-based
spectroscopic target selection procedure is expected to introduce a bias
against the most metal-rich RGB stars (\S\,\ref{sec:ddo51}).  We investigate
this by plotting $\rm[Fe/H]_{phot}$ versus $P_{\rm giant}$ in
Figure~\ref{fig:gprob_feh}.  There is no strong or obvious trend for the
primary and secondary spectroscopic targets ($P_{\rm giant}>0.5$) that form
the bulk of the sample.  The handful of lists~3--4 ``filler'' targets in our
sample with $P_{\rm giant}<0.5$ appear to have the same $\rm[Fe/H]_{phot}$
distribution as the rest of the sample---they are {\it not\/} particularly
metal rich.  This test will ultimately be carried out with a larger sample of
``filler'' targets from the full DEIMOS survey.  In summary, it appears that
$DDO51$-based pre-screening does not introduce any strong metallicity bias
over that introduced by $I$-band selection.

\subsubsection{Comparing Metallicity Distributions}\label{sec:feh_comp}

The metallicity distribution of RGB stars in M31's giant southern stream is
compared to that of its general halo in Figure~\ref{fig:feh}.  The top panel
shows radial velocity as a function of metallicity for 68~confirmed RGB
stars.  The portion below the dashed horizontal line is the velocity range
occupied by the stream, $v<-410$~km~s$^{-1}$.  It appears that the 47~stars
in this range are more skewed towards high metallicities than the 21~stars
outside it.  The thin solid histogram in the lower panel is the
$\rm[Fe/H]_{phot}$ distribution of stars in the stream's velocity range,
while the dashed histogram represents the non-stream general halo population.
We will show below that the stream contains a few stars brighter than the RGB
tip; $\rm[Fe/H]_{phot}$ estimates for such stars are unreliable because they
are based on a naive linear upward extrapolation of the RGB fiducials
(\S\,\ref{sec:feh_phot_method}).  We therefore recompute the stream's
metallicity distribution, this time excluding the seven~stars that lie above
the tips of the model RGB tracks (bold solid histogram).

The stream appears to be more metal-rich on average than the general halo,
$\rm\langle[Fe/H]\rangle=-0.51$ versus $-0.74$, and appears to have a smaller
metallicity spread, 0.25 versus 0.40~dex.  A two-sided Kolmogorov-Smirnov
test indicates that there is only a 5\% probability that the two~[Fe/H]
distributions are drawn from the same parent distribution---in other words,
the apparent difference between the stream and general halo metallicity
distributions is a $2\sigma$ effect.  Another difference between stream and
general halo populations is evident from the CMD [Fig.~\ref{fig:cmd}({\it
d\/})]: There are nine~M31 stars that lie near or above the RGB tip and all
nine are in the stream's radial velocity range (pentagons).  A couple of
these stars in particular are more than 0.5~mag above M31's RGB tip and yet
must be M31 members given their large negative radial velocities.  These are
best explained in terms of an intermediate-age asymptotic giant branch (AGB)
population in the stream.

Two~pieces of information suggest that the {\it true\/} mean metallicity of
the stream may be even higher, and the difference in mean metallicity between
the stream and general halo larger/more significant, than observed.  First,
if a sizeable fraction of the RGB stars in the stream are indeed of
intermediate age, then their photometric [Fe/H] estimates will need to be
corrected upward by about $+0.3$~dex.  Second, any bias resulting from our
photometric selection procedure will tend to deplete/truncate the high end of
the [Fe/H] distribution (\S\,\ref{sec:feh_bias}), the end where the
difference between the stream and general halo distributions appears to be
the greatest.

Even if the observed mean metallicity of the stream is taken at face value,
it points to a relatively high luminosity for the progenitor satellite galaxy
of the stream.  An empirical correlation between metallicity and luminosity
has been noted for dwarf satellite galaxies in the Local Group
\citep{mat98,gre99,dek03} which would indicate an absolute $B$-band magnitude
of $M_B\sim-17$ for the stream's progenitor, or a luminosity of
$L_B\approx10^9\,L_\odot$.  This is consistent with the lower limits on
progenitor luminosity derived from the stream's width and internal velocity
dispersion in Paper~II.

The mean [Fe/H] of $-0.7$~dex observed for the general halo population in
field~`a3', and its total range of 1.5~dex, are broadly consistent with
published studies of the metallicity distribution in other fields around the
M31 halo \citep*{mou86,dur94,dur01,dur04,ric96,hol96,rei98,sar01,bel03}.
However several factors make it difficult for us to carry out detailed
comparisons or to draw firm conclusions about real metallicity variations
from field to field.  First, our field~`a3' general halo [Fe/H] distribution
is based on only 21~RGB stars and therefore suffers from large Poisson
fluctuations.  Second, while our study like these others is based on
photometry of RGB stars, there are significant differences in terms of data
analysis techniques and associated systematic errors, sample definition,
contamination issues, and selection biases.  The RG02 study may be comparable
to ours since both are based on spectroscopy of RGB stars.  We note that RG02
found the halo to be significantly more metal poor, by $\gtrsim0.5$~dex in
the mean, with a tail in the [Fe/H] distribution extending down to
$\lesssim-2$~dex that is simply not seen in the present study [see their
Fig.~17({\it c\/})].  This may be yet another sign of substructure in M31's
halo, something that needs to be confirmed using larger, more uniformly
selected, and directly comparable samples.

\subsection{Spectroscopic Constraints on Metallicity}\label{sec:feh_spec}

The discussion of the chemical abundance distribution in M31's halo has so
far been based purely on photometric metallicity estimates.  In this section
we present a brief analysis of the Ca\,{\smcap ii} line strength in the
spectra of RGB stars.  The EW(Ca) measurement has a relatively large
uncertainty associated with it, compared to the random error in
$\rm[Fe/H]_{phot}$ say, so we prefer not to translate it into an estimate of
the spectroscopic metallicity $\rm[Fe/H]_{spec}$.  Instead the Ca\,{\smcap
ii} line strength is used as a point of comparison between photometric and
spectroscopic metallicity estimates.  We demonstrate that the two are in good
agreement.

\subsubsection{Predicted Calcium Line Strength}\label{sec:sigma_ca_pred}

The photometric properties of each star are used to predict the strength of
its Ca\,{\smcap ii} triplet.  First, the photometric metallicity estimate
$\rm[Fe/H]_{phot}$ (\S\,\ref{sec:feh_phot_method}) is taken to be the same as
the metallicity on the spectroscopic scale defined by \citet{car97},
$\rm[Fe/H]_{CG97}$.  Next, we use a well-established empirical calibration
relation, based on luminous RGB stars in Milky Way globular clusters, to
derive $\rm[Fe/H]_{CG97}$ from the Ca\,{\smcap ii} triplet
\citep*{rut97a,rut97b}:
\begin{equation}
{\rm[Fe/H]_{CG97}~=~-2.66\,+\,0.42\>[\Sigma{Ca}}\,-\,0.64(V_{\rm HB}-V)]
\label{eqn:feh_cg97}
\end{equation}
where $\rm\Sigma{Ca}$ is the weighted sum of the EWs, in units of \AA, of the
three~lines comprising the Ca\,{\smcap ii} triplet:
\begin{equation}
\rm\Sigma{Ca}\,\equiv\,0.5\>EW(\lambda8498\AA)\,+\,1.0\>EW(\lambda8542\>\AA)
\,+\,0.6\>EW(\lambda8662\>\AA)
\label{eqn:sigma_ca}
\end{equation}
and the luminosity-based correction for the effect of surface gravity is made
relative to the apparent magnitude of M31's horizontal branch: $V_{\rm
HB}=25.17$ \citep{hol96}.  Inverting equation~(\ref{eqn:feh_cg97}), the {\it
predicted\/} Ca\,{\smcap ii} line strength is defined to be:
\begin{equation}
{\rm\Sigma{Ca}_{pred}~\equiv~6.33\,+\,2.38\>[Fe/H]_{phot}}\,+\,0.64(V_{\rm
HB}-V)
\label{eqn:sigma_ca_pred}
\end{equation}
Since this relation is based on Galactic globular cluster RGB stars, we are
making the implicit assumption that M31 RGB stars are comparably old
($t\gtrsim10$~Gyr) and, more importantly, alpha-enhanced to the same degree
($\rm[\alpha/Fe]=+0.3~dex$).  As discussed in \S\,\ref{sec:coadd_spec} below,
RGB stars are grouped according to the $\rm\Sigma{Ca}_{pred}$ parameter for
the purpose of coadding spectra.

\subsubsection{Stellar Absorption Features and Coadded
Spectra}\label{sec:coadd_spec}

The normalized 1D spectra shown in Figure~\ref{fig:spec_indiv} are arranged
in order of the stars' $(V-I)_0$ color becoming redder upwards.  The onset of
the TiO band seems to occur at $(V-I)_0\sim1.8$ which corresponds to $T_{\rm
eff}=4000\,$K \citep{alo99}.  As might be expected, there is a good
correlation between the observed broad-band color and the strength of the TiO
bands for stars that are redder (cooler) than this.  Other features, such as
the Ca\,{\smcap ii} triplet and occasionally the Na\,{\smcap i} doublet, are
also visible.  It should be noted however that the 1D spectra shown in
Figure~\ref{fig:spec_indiv} represent the best third of our sample.
Unfortunately the spectral S/N ratio is not high enough to support a detailed
star-by-star abundance analysis of all RGB stars in our sample.

We have decided therefore to coadd the 92~spectra (including 8~duplicate
observations) in groups of about a dozen to improve the S/N.  The spectra are
grouped according to the expected Ca\,{\smcap ii} triplet absorption line
strengths of the stars, $\rm\Sigma{Ca}_{pred}$.  Two~different grouping
schemes are tried.  In the first scheme, no distinction is made between
stream versus general halo populations and the entire set of RGB stars are
simply divided into five~bins by line strength.  In the second scheme, the
47~RGB stars in the stream's radial velocity range are divided into four~bins
and the 21~general halo RGB stars outside this range into two~bins.  In both
schemes, the 16~Galactic dwarf stars are placed in a bin by themselves.  All
spectra are shifted to zero velocity (rest frame) and then combined with
inverse-variance weighting.  The resulting coadded spectra are then smoothed
with a weighted boxcar of width 5~pixels ($\rm1.7\AA$), comparable to the
instrumental resolution.

Figure~\ref{fig:spec_coadd} shows the six~coadded spectra using the first
grouping scheme.  The lowest coadded spectrum is that of Milky Way dwarfs and
the next five are M31 RGB star coadded spectra with predicted line strength
increasing upwards.  The top spectrum (bold line) is a model RGB spectrum
from \citet{sch99} computed using $T_{\rm eff}=4000\,$K, $\log(g)=1.5$, and
$\rm[Fe/H]=-0.3$.  It is merely used to illustrate/identify usable absorption
features in the far red region of the spectrum---e.g.,~Ca, Fe, Mg, Ti, and V
lines for RGB stars, and these plus the surface-gravity-sensitive Na\,{\smcap
i} doublet for dwarfs.  A detailed quantitative treatment of absorption lines
is postponed until a future paper; only the Ca\,{\smcap ii} line strengths
are discussed briefly below.  We have therefore not made any attempt to fine
tune the match between model and data in terms of line strength or spectral
resolution.

\subsubsection{Measured Calcium Line Strength and Some Sanity
Checks}\label{sec:sigma_ca_comp}

\begin{figure}
\centerline{\epsfxsize=6.2in \epsfysize=6.2in
\epsfbox{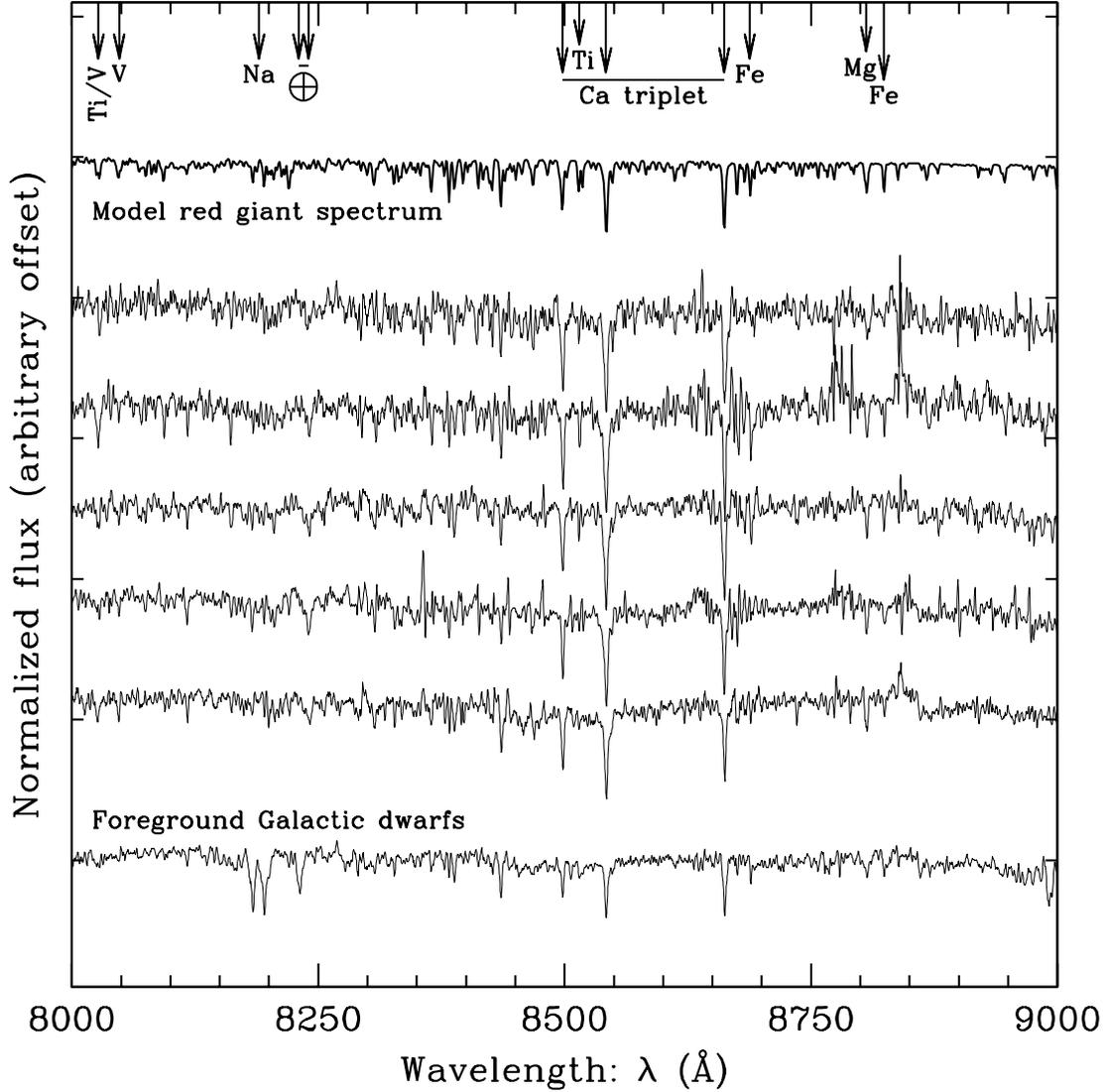}}
\caption{\label{fig:spec_coadd}{
Montage of coadded DEIMOS spectra (thin solid lines) showing the region
around the Ca\,{\smcap ii} triplet, normalized and shifted to rest-frame
wavelength and smoothed with a 1.7\AA\ weighted boxcar.  The lowest spectrum
is a coadd of 16~foreground Galactic dwarf stars.  The next five~spectra are
coadds of about a dozen M31 RGB stars each, grouped and ordered by {\it
predicted\/} Ca\,{\smcap ii} line strength (increasing upward) as estimated
from the CMD-based photometric metallicity and luminosity
(\S\,\ref{sec:sigma_ca_pred}).  The bold line at the top is a model red giant
spectrum with $T_{\rm eff}=4000\,$K, $\log(g)=1.5$, and $\rm[Fe/H]=-0.3$ from
\citet{sch99}.  A few prominent spectral features of RGB stars are
identified along with the $\rm\lambda8190\AA$ Na\,{\smcap i} doublet, which
is strong in dwarfs.  Because we have corrected the M31 RGB and Galactic
dwarf spectra to zero velocity, the $\rm\lambda8228\AA$ telluric feature is
Doppler shifted to $\sim\rm\lambda8231\AA$ and $\sim\rm\lambda8240\AA$,
respectively.
}}
\end{figure}

The EWs of the Ca\,{\smcap ii} lines are measured in all coadded spectra
using an 18\AA-window for each line.  The continuum level (close to unity,
since these are normalized spectra) is measured on either side of each
Ca\,{\smcap ii} line using windows of the same width while avoiding regions
of the spectrum known to contain other strong lines (see RGB model spectrum
in Fig.~\ref{fig:spec_coadd}).  For each spectrum the weighted sum of the EWs
of the three~Ca\,{\smcap ii} lines, $\rm\Sigma{Ca}_{meas}$, is computed
according to equation~(\ref{eqn:sigma_ca}).

\begin{figure}
\centerline{\epsfxsize=6in \epsfysize=6in
\epsfbox{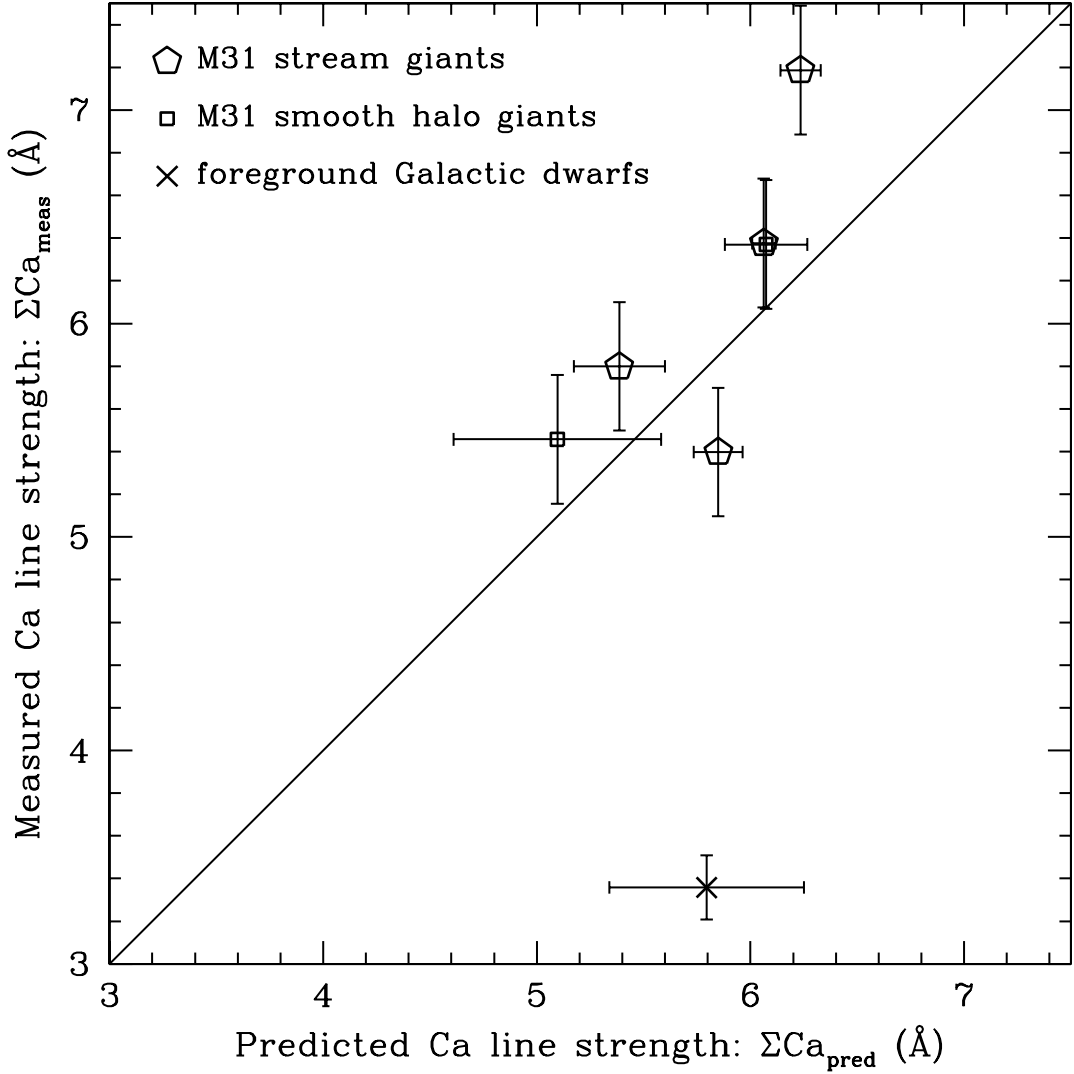}}
\caption{\label{fig:ca_pred_meas}{
Measured equivalent width of the Ca\,{\smcap ii} triplet lines in coadded
spectra plotted as a function of the predicted EW.  The EW $\rm\Sigma{Ca}$ is
defined to be the weighted sum of the EWs of the three~lines
(\S\,\ref{sec:sigma_ca_pred}).  The $\rm\Sigma{Ca}_{pred}$ value for each
star is derived from its CMD/RGB fiducial-based metallicity estimate
$\rm[Fe/H]_{phot}$ and stellar luminosity, as described in the text.  Stars
are placed in groups of a dozen or more according to $\rm\Sigma{Ca}_{pred}$
and their spectra coadded.  M31 RGB stars in the giant southern stream
(pentagons) and general halo (squares) show reasonable agreement between
predicted and measured EWs, with the exception of the strongest-lined (most
metal rich) stream stars for which $\rm\Sigma{Ca}_{meas}>\Sigma{Ca}_{pred}$.
The $\rm\Sigma{Ca}_{pred}$ calculation makes no sense for foreground Galactic
dwarf stars so it is no surprise that they fall well off the one-to-one
relation (cross).
}}
\end{figure}

Figure~\ref{fig:ca_pred_meas} compares the measured versus predicted combined
Ca\,{\smcap ii} EWs for the seven~coadded spectra from the second grouping
scheme defined above.  It is reassuring to see that the M31 RGB coadds, both
stream and general halo stars (pentagons and squares, respectively), lie
close to the one-to-one line.  There is a slight mismatch for the most
strong-lined (metal-rich) stream RGB stars.  If these happen to be
intermediate-age stars---i.e.,~in the $\approx2$--6~Gyr range instead of the
12.6~Gyr age that is assumed in fitting model RGB fiducials---their
photometric metallicity estimates would be biased low by $-0.3$~dex
(\S\,\ref{sec:feh_phot_method}).  Thus the $\rm\Sigma{Ca}_{pred}$ values
would be biased low by 0.7\,\AA (eq.~[\ref{eqn:sigma_ca_pred}]) and would,
for the most part, explain the observed offset from the one-to-one line.
This assumes that intermediate-age RGB stars follow the same
$\rm\Sigma{Ca}\rightarrow[Fe/H]_{CG97}$ calibration relation as old RGB stars
(eq.~[\ref{eqn:feh_cg97}]), which is yet to be verified.

Eight of the nine~potential intermediate-age AGB stars (near/above the RGB
tip in the CMD) are in the first bin for spectral coadds of stream stars in
Figure~\ref{fig:ca_pred_meas}---i.e.,~they have among the lowest
$\rm\Sigma{Ca}_{pred}$ values and are expected to have relatively weak
Ca\,{\smcap ii} lines.  This bin also contains 4~normal metal-poor RGB stars
located well below the RGB tip.  The coadded spectrum for this group of stars
appears to have a ``normal'' Ca\,{\smcap ii} triplet strength, in that it
lies close to the $\rm\Sigma{Ca}_{meas}=\Sigma{Ca}_{pred}$ line, but it is
not clear how to interpret this.  If the stars above/near the RGB tip are
indeed intermediate-age AGB stars: (i)~$\rm[Fe/H]_{phot}$ estimates are bound
to be inaccurate as they are based on an arbitrary extrapolation of the RGB
fiducials (\S\,\ref{sec:feh_phot_method}); and (ii)~the empirical
$\rm\Sigma{Ca}\rightarrow[Fe/H]_{CG97}$ calibration relation
(eq.~[\ref{eqn:feh_cg97}]), which is based on RGB stars, is likely to be off
for AGB stars.  It is conceivable that these two~errors are somehow
cancelling each other out; a detailed investigation of these issues is beyond
the scope of this paper.

Judging from their location on the CMD and the agreement between
$\rm\Sigma{Ca}_{pred}$ and $\rm\Sigma{Ca}_{meas}$, it appears that about
two-thirds of the stars in our M31 RGB sample are old: This includes the
two~bins containing general halo members and bins~2 and 3 from the stream
population in Figure~\ref{fig:ca_pred_meas}.  If the nine~stream stars
located above the RGB tip in the CMD (most are in stream bin~1) and the
dozen strong-lined stream RGB stars located above/to the left of the
$\rm\Sigma{Ca}_{meas}=\Sigma{Ca}_{pred}$ line (stream bin~4) all turn out to
have ages $t\lesssim8$~Gyr as suspected, it would imply that about 30\% of
the overall M31 RGB population in field~`a3' is of intermediate age.

In a recent deep {\it Hubble Space Telescope\/} study of M31 main-sequence
turnoff stars in an inner halo field ($R=7$~kpc), \citet{bro03} found a
surprisingly high fraction of intermediate-age stars, $\approx30$\%.  It has
been suggested that the orbit of the giant southern stream might wrap around
and intersect the \citeauthor{bro03} field and that this might be responsible
for the high intermediate-age fraction seen there.  If this explanation were
correct, and taking our estimate of the field~`a3' intermediate-age fraction
literally, the stream would stand out against the smooth halo population in
the inner $R=7$~kpc with the same 2:1 contrast as it does in our remote
field~`a3'.  No such wrap-around portion of the stream is visible in the
\citet{fer02} star-count map.

In contrast to the RGB stars, the foreground Galactic dwarf stars lie well
below the $\rm\Sigma{Ca}_{meas}=\Sigma{Ca}_{pred}$ line in
Figure~\ref{fig:ca_pred_meas}.  The same effect was noted by RG02.  This is
yet another feature that can be used to distinguish between foreground dwarf
star contaminants and M31 RGB stars (\S\,\ref{sec:reject_Ca_triplet}).  It is
evident from Figure~\ref{fig:spec_coadd} that the coadded dwarf star spectrum
is significantly less noisy than the M31 RGB star coadds: This is because the
former are $\Delta{I}\approx1$~mag brighter on average [compare panels~({\it
c\/}) and ({\it d\/}) of Fig.~\ref{fig:cmd}] and there are 16~stars in the
former coadd versus 12 or 13 in the latter.  The error in
$\rm\Sigma{Ca}_{pred}$ for the dwarfs is estimated to be 50\% that of the M31
RGB stars.  Thus, the offset of the dwarfs from the one-to-one line is highly
significant.

\section{Summary}\label{sec:summary}

The following are the main points of this paper:

\begin{itemize}
\item[$\bullet$]{We are using the DEIMOS spectrograph on the Keck~II 10-meter
telescope to carry out a moderate-resolution ($R\lesssim7000$) spectroscopic
survey of a large sample of RGB stars in the outer halo of M31.  This is the
first paper from that survey and describes data from three~DEIMOS slitmasks
in field~`a3', located on the giant southern debris stream discovered by
\citet{iba01} at a projected distance of 31~kpc from the center of M31.  The
field~`a3' data presented here ($\gtrsim200$~spectroscopic targets) represent
about a quarter of the DEIMOS survey data obtained to date.}

\item[$\bullet$]{Spectroscopic targets were selected using intermediate-band
$DDO51$ and Washington $M$ and $T_2$ photometry by discriminating between RGB
stars and foreground Galactic dwarfs on the basis of surface gravity.  The
method has proved to increase the yield of bona fide M31 RGB stars in our
sample.}

\item[$\bullet$]{A sample of 68~definite M31 RGB stars is isolated.  Careful
attention is paid to the removal of sample contaminants, both background
galaxies and especially foreground Milky Way dwarf stars.  The latter are
identified using a combination of data/methods: radial velocity, broad-band
color-magnitude information, $DDO51$ photometry, Ca\,{\smcap ii} triplet line
strength, and, most importantly, Na\,{\smcap i} doublet line strength.}

\item[$\bullet$]{About two-thirds of the M31 RGB stars in our field appear to
be members of its giant southern stream while the rest belong to the general
halo population.}

\item[$\bullet$]{The mean heliocentric radial velocity of the stream in
field~`a3' is $-458$~km~s$^{-1}$ which translates to $-158$~km~s$^{-1}$ with
respect to the systemic velocity of M31.  The stream has a relatively low
internal line-of-sight velocity dispersion: $15_{-15}^{+8}$~km~s$^{-1}$ (90\%
confidence limits from a maximum likelihood analysis).  The interpretation of
these and other data on the kinematics and three-dimensional structure of the
stream, in the context of possible orbits and progenitor properties, is
presented in the companion paper by \citet[][Paper~II]{fon04}.}

\item[$\bullet$]{The rms metallicity spread of M31's giant southern stream is
0.25~dex and its mean metallicity is $\rm\langle[Fe/H]\rangle=-0.51$,
possibly higher if one corrects for selection bias against the highest
metallicity RGB stars.  This is indicative of a fairly luminous progenitor
satellite galaxy.  The photometric and spectroscopic metallicity estimates
are in good agreement with each other for the majority of RGB stars in our
sample.}

\item[$\bullet$]{The most metal rich RGB stars in the stream have
$\rm[Fe/H]_{spec}>[Fe/H]_{phot}$ (anomalously strong Ca\,{\smcap ii} lines).
The stream also contains a few stars that lie above the RGB tip in the CMD.
Both findings suggest that the stream contains a non-negligible fraction of
intermediate-age stars.}

\item[$\bullet$]{The general halo population in field~`a3' has a mean
metallicity of $\rm\langle[Fe/H]\rangle=-0.74$ and thus appears to he more
metal poor than the stream on average.  The halo component has a broad
metallicity distribution spanning about 1.5~dex.}

\item[$\bullet$]{There is a hint of halo substructure in M31 based on the
radial velocity (and possibly metallicity) distribution of the general halo
RGB population in this field.}
\end{itemize}

\acknowledgments

We are grateful to Sandy Faber and the DEIMOS team for building an
outstanding instrument and for extensive help and guidance during its first
observing season.  We also thank Alison Coil, Drew Phillips, and Greg Wirth
for observing field~`a3' masks~\#2 and \#3 on our behalf, Drew Phillips for
help with slitmask designs, Ricardo Schiavon for providing a model RGB
spectrum and expert advice on spectral features, Jeff Lewis and Matt Radovan
in the Lick instrument shops for their careful and timely fabrication of the
masks, and the DEEP2 team for allowing us use of the {\tt spec1d}/{\tt zspec}
software.  The {\tt spec2d} data reduction pipeline for DEIMOS was developed
at UC Berkeley with support from NSF grant AST-0071048.  P.G.\ acknowledges
support from NSF grant AST-0307966 and a Special Research Grant from UCSC.
R.M.R.\ and D.B.R.'s contributions were supported by NSF grant AST-0307931.
S.R.M., J.C.O., and R.J.P.\ acknowledge funding by NSF grants AST-0307842 and
AST-0307851, NASA/JPL contract 1228235, the David and Lucile Packard
Foundation, and The F.~H.~Levinson Fund of the Peninsula Community
Foundation.  M.G.\ is supported by NASA through Hubble Fellowship grant
HF-01159.01-A awarded by the Space Telescope Science Institute, which is
operated by the Association of Universities for Research in Astronomy, Inc.,
under NASA contract NAS~5-26555.  K.V.J.'s contribution was supported through
NASA grant NAG5-9064 and NSF CAREER award AST-0133617.

\clearpage

\end{document}